\pdfoutput=1
\documentclass[12pt]{article}
\usepackage[utf8]{inputenc}
\usepackage{tabularx}
\usepackage{longtable}

\usepackage[left=2cm,right=2cm,top=2cm,bottom=2cm]{geometry}   
\usepackage[small]{caption}
\usepackage{relsize}
\usepackage[table,svgnames,dvipsnames]{xcolor}
\usepackage{enumitem}
\usepackage{bbold}
\usepackage[toc,page]{appendix}
\usepackage{listings}   
\usepackage[pdftex]{graphicx}
\usepackage{latexsym,amsmath,amsfonts,amssymb}
\usepackage[numbers,sort&compress]{natbib}
\usepackage{url}
\usepackage[colorlinks=true
  ,urlcolor=blue
  ,anchorcolor=blue
  ,citecolor=blue
  ,filecolor=blue
  ,menucolor=blue
  ,linktocpage=true
  ,pdfproducer=medialab
  ,pdfa=true
  ,linktoc=all
]{hyperref}  

\usepackage{amssymb} 
\usepackage{amsmath}                    			% Mathematical things
\usepackage{amssymb}                    			% More mathematics
\usepackage{mathtools}                  			% equal explanation
\usepackage{latexsym}                   			% Some symbols
\usepackage{bbold}                      			% Real, Natural, Complex... \mathbb{}
\usepackage{braket}                     			% Dirac Notation
\usepackage{graphicx}                               % Inserting images              % Multiple images
\usepackage{subcaption}
\usepackage{wrapfig}                    			% Text around figure
\usepackage{sidecap}                    			% Minipage (position image-equations)
\usepackage{physics}                    			% Physics language
\usepackage{color}                      			% Colored text
\usepackage[dvipsnames]{xcolor}         		    % More colors
\usepackage{soul}                       			% \ul{} underline multiple lines       % Refs in text
\usepackage{array}                      			% Multirow in tables
\usepackage{enumitem}                   			% Enumerate, items
\usepackage{tensor}                     			% Tensors
\usepackage{cancel}                     			% Cancel terms in equations
\usepackage{comment}                    			% Comment whole paragraph
\usepackage{empheq}                                 % Equations. Necessary for fbox

\usepackage{tikz}
\usepackage{bbm}
\usepackage{float}
\usepackage{pdfscreen}                              % For slides outside beamer
\usepackage[display]{texpower}                      % For slides outside beamer
\usepackage{pifont}% http://ctan.org/pkg/pifont     % Nice checkmark and crossmark
\usepackage[thinlines]{easytable}
\usetikzlibrary{decorations.pathreplacing}
\usetikzlibrary{decorations.pathmorphing}
\usepackage{indentfirst}                            % Indent first paragraph of sections
\usepackage{csquotes}                               % Quotations
\MakeOuterQuote{"}                                  % Quotations

\numberwithin{equation}{section}

\begin{document}
\begin{center}
 
\centerline{\LARGE {\bf Hot wormholes and chaos dynamics  }}\vspace{0.4cm}
\centerline{\LARGE {\bf in a two-coupled SYK model}}

\vspace{8mm}
\hypersetup{linkcolor=black}
\renewcommand\thefootnote{\mbox{$\fnsymbol{footnote}$}}
Mart\'i Berenguer\footnote{marti.berenguer.mimo@usc.es},
Javier Mas\footnote{javier.mas@usc.es},
Juan Santos-Su\'arez\footnote{juansantos.suarez@usc.es}
and  Alfonso V. Ramallo\footnote{alfonso.ramallo@usc.es}

\vspace{4mm}

{\small \sl Departamento de  F\'isica de Part\'iculas} \\
{\small \sl Universidade de Santiago de Compostela} \\
{\small \sl and} \\
{\small \sl Instituto Galego de F\'isica de Altas Enerx\'ias (IGFAE)} \\
{\small \sl E-15782 Santiago de Compostela, Spain} 
\vskip 0.2cm

\end{center}

\vspace{8mm}
\numberwithin{equation}{section}
\setcounter{footnote}{0}
\renewcommand\thefootnote{\mbox{\arabic{footnote}}}

\begin{abstract}

We study the dynamics of chaos across the phase transition in a 2-coupled Sachdev-Ye-Kitaev (SYK) model, with a focus on the unstable "hot wormhole" phase. Using the Schwinger-Keldysh formalism, we employ two non-equilibrium protocols that allow access to this phase, which is inaccessible through equilibrium simulations: one involves cooling the system via a coupling to a thermal bath, while in the other we periodically drive the coupling parameter between the two sides. We numerically compute the Lyapunov exponents of the hot wormhole for the two cases. Our results uncover a rich structure within this phase, including both thermal and non-thermal solutions. These behaviors are analyzed in detail, with partial insights provided by the Schwarzian approximation, which captures certain but not all aspects of the observed dynamics.

\end{abstract}

\clearpage

\hypersetup{linkcolor=black}
\tableofcontents
\hypersetup{linkcolor=blue}

\section{Introduction}

The study of quantum chaos and information scrambling has emerged as a cornerstone of modern theoretical physics, providing profound insights into the behavior of quantum systems and their gravitational duals. Quantum chaos, characterized by a sensitive dependence on initial conditions, manifests in complex many-body systems as the rapid delocalization of quantum information. In the large-$N$ regime, a key signature of this chaotic behavior is the exponential growth of out-of-time-ordered correlators (OTOCs), characterized by the Lyapunov exponent, $\lambda_L$. Central to this area of research is the quest to understand how quantum information propagates over time in systems governed by strongly interacting degrees of freedom. This phenomenon has far-reaching implications for our understanding of black holes and the fundamental nature of spacetime \cite{Hayden:2007cs, Sekino:2008he,Shenker:2013pqa}.

The Sachdev-Ye-Kitaev (SYK) model \cite{Sachdev_1993, KitaevTalk1, KitaevTalk2, MaldaStanford} has emerged as a powerful and controllable framework for testing these ideas. Comprising $N$ Majorana fermions with random all-to-all interactions, the SYK model is exactly solvable in the large-$N$ limit and exhibits conformal symmetry at low energies. In this regime, it has been shown to exhibit maximal chaos, saturating the bound proposed in \cite{Maldacena:2015waa}. Furthermore, the SYK model is closely related to Jackiw-Teitelboim (JT) gravity, a two-dimensional gravitational theory \cite{Maldacena:2016upp}. This remarkable connection has spurred extensive research exploring the deep link between strongly coupled quantum systems and gravitational physics.

A particularly interesting generalization was proposed in \cite{MaldaQi}, involving two identical copies of the SYK model coupled through a bilinear interaction. This setup offers a tractable framework for exploring phases of matter with distinct gravitational duals. The model exhibits a Hawking-Page-like transition between a gapless, high-temperature phase and a gapped, low-temperature phase. It has been argued that these phases correspond to duals of a two-sided AdS$_2$ black hole and global AdS$_2$ spacetime, respectively. In the low-temperature phase, the two boundaries of the AdS$_2$ spacetime are causally connected, representing a traversable wormhole geometry created by negative null energy resulting from the coupling between the two boundaries \cite{Gao_2017, Maldacena_2017}. Notably, this traversable wormhole serves as a laboratory for studying quantum teleportation protocols: it has been demonstrated that a small perturbation applied to one SYK system can effectively teleport quantum information to the other, exploiting the wormhole's geometric connectivity \cite{Gao:2019nyj, Brown_2023, Nezami_2023}. This remarkable phenomenon has significantly advanced our understanding of the interplay between quantum information theory and gravitational physics.

The Lyapunov exponent for the two stable phases of the model has been previously computed in \cite{Nosaka_2020}. However, the model also exhibits an unstable phase, commonly referred to as the "hot wormhole" phase, which is inaccessible via equilibrium simulations in the canonical ensemble \cite{MaldaQi}. In this paper, we leverage the Schwinger-Keldysh (SK) formalism \cite{Kamenev_2011} to dynamically probe the hot wormhole phase and compute its chaos exponents. This is achieved through two distinct out-of-equilibrium protocols.

The first protocol, proposed in \cite{MaldaMilekhin}, involves starting from a high-temperature (black hole) solution and cooling down the system to a wormhole solution by coupling it to a cold bath. This cooling process enables a quasi-static evolution that transitions through the unstable phase. The second method, introduced in \cite{Berenguer_2024}, involves periodically driving the coupling between the two sides over time. This approach injects energy into the system, allowing access to the hot wormhole phase without relying on an external bath, which preserves the intrinsic dynamics of the original system. By driving the coupling parameter, we uncover a richer structure of potential evolution endpoints within the unstable phase. These findings are confirmed by a third non-equilibrium protocol that combines the two previous ones. We also perform a qualitative analysis using the Schwarzian approximation, shedding light on the dynamics of this exotic phase.

Exploring the dynamics of the model beyond equilibrium remains an open and intriguing area of research. Recent studies have investigated the model's out-of-equilibrium behavior \cite{MaldaMilekhin, Zhou:2020kxb, Zhang_2021, Berenguer_2024}, where processes such as cooling or heating can probe the stability of the wormhole and uncover additional features of the intermediate "hot wormhole" phase. These explorations are particularly significant for addressing fundamental questions about thermalization, quantum chaos, and the role of entanglement in holographic dualities. The model thereby offers a robust framework for studying not only static geometries but also the time-dependent processes that drive transitions between phases with different spacetime connectivity.

\begin{comment}
The protocol was introduced in \cite{Berenguer_2024}, and consists of doing the inverse process: the system begins in a WH solution, and then the coupling $\mu$ is driven periodically as  $\mu(t)=\mu(1+a\sin \Omega t)$ in order to inject energy in the system. If after some finite time the driving is turned off, the system will eventually reach an equilibrium configuration of higher energy. If the supplied energy is smaller than the energy gap between the two stable phases, the system will equilibrate to a hot wormhole solution, as it was checked numerically in \cite{Berenguer_2024}, where the process was repeated for different durations of the driving, reconstructing (part of)\footnote{See Section \ref{sec:FloquetSYK}.} the microcanonical equilibrium curve of the phase diagram.
\end{comment}

The paper is organized as follows: In Section \ref{sec:model}, we review the 2-coupled SYK model \cite{MaldaQi} and the cooling protocol introduced in \cite{MaldaMilekhin}. Section \ref{sec:HWchaos} focuses on the analysis of the hot wormhole solutions obtained through the two distinct driving protocols. We compare these solutions with the two stable phases of the model and numerically compute their Lyapunov exponents. In Section \ref{sec:HotWHColdBH}, we conduct a deeper investigation of the unstable region in the phase diagram and use the Schwarzian approximation to provide a qualitative explanation of the observed results. Finally, in Section \ref{sec:conclusions}, we present a summary of our findings and discuss potential directions for future research.

\section{The model}\label{sec:model}
We consider the two-coupled SYK model introduced in \cite{MaldaQi}, which consists of two identical SYK systems, each containing $N$ Majorana fermions with all-to-all random interactions of order $q$. In this work, we focus on the case $q=4$. The two systems are coupled through a bilinear interaction that pairs fermions from each system. The Hamiltonian is given by
\begin{equation}   
    H=\sum_{a=L,R}\frac{1}{4!}\sum_{ijkl}J_{ijkl}\chi_a^i\chi_a^j\chi_a^k\chi_a^l+i\mu\sum_j \chi_L^j\chi_R^j~,
    \label{eq:Hamiltonian}
\end{equation}
where $\chi_a^i$, with $i=1,...,N$ and $a=L,R$, are Majorana fermions of the left/right SYK model, respectively, satisfying the usual anticommutation relations $\left\{\chi_a^i,\chi_b^j\right\}=\delta^{ij}\delta_{ab}$. The couplings $J_{ijkl}$ are real constants drawn from a gaussian distribution with zero mean and variance given by
\begin{equation}
\overline{J_{ijkl}^2}=\frac{3! J^2}{N^3} ~.
\label{eq:varianceJ}
\end{equation}

After averaging over the random couplings and introducing  bi-local fields
\begin{equation}
    G_{ab}(\tau,\tau')=\frac{1}{N}\sum_j\langle \chi_a^j(\tau)\chi_b^j(\tau')\rangle~,
\end{equation}
the averaged partition function can be written in terms of a Euclidean effective action as follows \cite{MaldaQi,Berenguer_2024}
\begin{equation}
    \overline{Z}=\int\mathcal{D}G_{ab}\mathcal{D}\Sigma_{ab}~e^{-S_{eff}[G,\Sigma]}~,
\end{equation}
with
\begin{align}
\begin{split}
    \frac{S_{eff}[G,\Sigma]}{N}=&~-\frac{1}{2}\log\det \left(\delta(\tau-\tau')\left(\delta_{ab}\partial_\tau-\mu \sigma_{ab}^y\right)-\Sigma_{ab}(\tau,\tau')\right)\\
    &+\frac{1}{2}\int d\tau d\tau'\sum_{a,b}\Sigma_{ab}(\tau,\tau')G_{ab}(\tau,\tau')-\frac{J^2}{8}\sum_{a,b}\int d\tau d\tau' \left[G_{ab}(\tau,\tau')\right]^4~,
\label{eq:Seffeuclidean}
\end{split}
\end{align}
and $\sigma_{LL}^y=\sigma_{RR}^y=0$, $\sigma_{LR}^y=-\sigma_{RL}^y=-i$.

In the large $N$ limit the dynamics of the system are governed by the saddle point equations 
\begin{align}
\begin{split}
    G_{LL}(i\omega_n)&=-\frac{i\omega_n+\Sigma_{LL}}{(i\omega_n+\Sigma_{LL})^2+(i\mu-\Sigma_{LR})^2}\\[10pt]
    G_{LR}(i\omega_n)&=-\frac{i\mu-\Sigma_{LR}}{(i\omega_n+\Sigma_{LL})^2+(i\mu-\Sigma_{LR})^2}~,
\label{eq:SDequilibrium}
\end{split}
\end{align}
where the self-energies $\Sigma_{ab}$ are given by
\begin{equation}
    \Sigma_{ab}(\tau)=J^2 G_{ab}(\tau)^3~.    
    \label{eq:Sigmaeq}
\end{equation}

The physics of the coupled model is characterized by two stable phases in the $(T, \mu)$ parameter space, with $J = 1$ set throughout this work. At low temperatures and small coupling $\mu$, the system exhibits a gapped phase identified as a traversable wormhole geometry. This phase corresponds to a global AdS$_2$ spacetime, where the two boundaries are smoothly connected through the interior. The wormhole arises due to the bilinear coupling, which induces strong correlations between the two SYK systems. Notably, the ground state of the model closely resembles the thermofield double (TFD) state of the single-sided system \cite{MaldaQi}. In this regime, the dynamics of the coupled system are governed by the same reparametrization mode that underlies the nearly-conformal physics of the SYK model. At these energy scales, the system exhibits sharp revival oscillations in fermion correlation functions, with a frequency proportional to $\mu^{2/3}$ \cite{Plugge_2020, Haenel_2021, Zhang_2021}, signaling the transfer of information between the two SYK systems. In the gravitational dual description, this information transfer occurs via the wormhole geometry.

At high temperatures or sufficiently large coupling $\mu$, the system transitions into a phase where the two SYK models behave approximately as decoupled systems. From a gravitational perspective, this phase corresponds to two disconnected AdS$_2$ black holes. The entanglement between the two boundaries is significantly reduced in this phase, and the wormhole geometry breaks down. This transition is analogous to the Hawking-Page transition between thermal AdS and AdS black hole geometries in higher dimensions.

\begin{figure}
    \centering
    \begin{subfigure}[t]{0.5\textwidth}
        \centering
        \includegraphics[width=\textwidth]{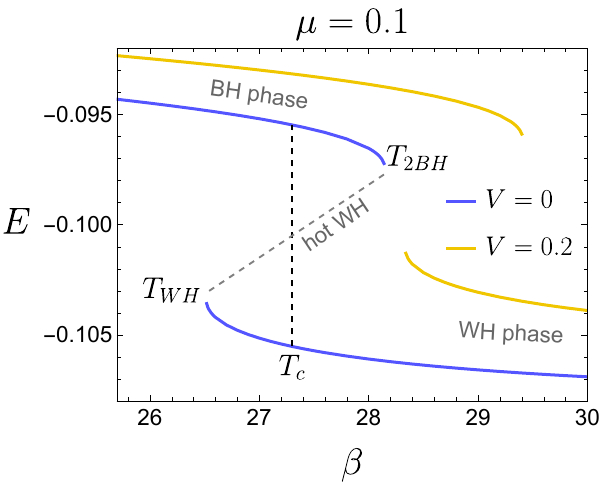}
        \label{sfig:phasediagE}
    \end{subfigure}
    \hspace{0.02\textwidth} % Add a little space between the subfigures
    \begin{subfigure}[t]{0.36\textwidth}
        \centering
        \vspace{-5.5cm} % Adjust this value to vertically center the second figure
        \includegraphics[width=\textwidth]{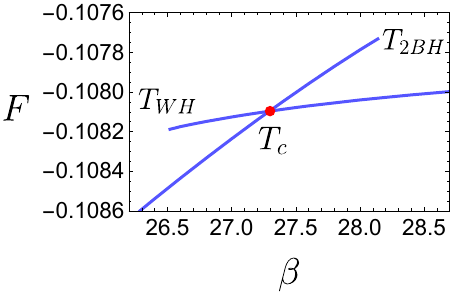}
        \label{sfig:phasediagF}
    \end{subfigure}
    \caption{Left: Phase diagram of the coupled model for $\mu=0.1$ (blue lines). In yellow we show the phase diagram of the model coupled to a cold bath, with coupling strength $V$, discussed further in Section \ref{sec:WHformation}. Right: The transition inverse temperature $\beta_c\sim 27.3$ is obtained by comparing the free energies in the Euclidean formalism, computed as the on-shell effective action \eqref{eq:Seffeuclidean}. The red dot corresponds to the vertical line in the left figure.}
    \label{fig:phasediag}
\end{figure}

Connecting these two phases is an unstable intermediate phase, referred to as the "hot wormhole" \cite{MaldaQi, MaldaMilekhin}. While this phase cannot be observed in equilibrium canonical simulations due to its instability, it has been conjectured to be stable in the microcanonical ensemble \cite{MaldaQi}. The phase diagram for $\mu=0.1$ is shown in Fig.~\ref{fig:phasediag}. The unstable phase can be analyzed analytically in the large-$q$ limit. In this limit, as the system transitions between phases, the energy evolves smoothly, suggesting that, although the geometry undergoes a discrete change, the underlying quantum dynamics remain continuous.

This observation is particularly relevant for cooling processes, where the system can dynamically relax from a high-energy state with decoupled black holes into the low-energy wormhole ground state. As demonstrated in \cite{MaldaMilekhin}, coupling the system to a cold bath allows it to explore the unstable phase during the transition. Notably, for $q=4$, the smooth transition between the black hole and wormhole phases persists, indicating that the system effectively traverses the unstable phase as part of the relaxation process.

\iffalse
In the next section, we will discuss how the coupled model can be evolved in real time, including the process of cooling from a black hole to a wormhole configuration, following the protocol introduced in \cite{MaldaMilekhin}.
\fi

\subsection{Wormhole formation in real time}\label{sec:WHformation}

In this section, we discuss the real-time dynamics of the 2-coupled SYK model after it is suddenly coupled to a cold bath, along the lines given in \cite{MaldaMilekhin}.  

The system begins in a high-temperature (BH) state at $T>T_{2BH}$, where the dynamics of the two SYK models are mostly decoupled. At $t=0$, we couple the system to a cold thermal bath with temperature $T_B < T_{WH}$, so that the system evolves towards an equilibrium wormhole configuration at $T=T_B$.
The numerical results suggest that the system approximately follows the microcanonical equilibrium curve during the evolution, thereby passing through the unstable hot wormhole phase of Fig.~\ref{fig:phasediag}.

To model the bath, we consider two independent SYK systems, each composed of $M$ fermions, denoted by $\psi^i$ and $\tilde{\psi}^i$. The distribution of random couplings is gaussian as in  \eqref{eq:varianceJ} with the replacement $N\rightarrow M$, and its variance denoted by $J_B$. Finally, each  SYK-bath term is coupled to a corresponding SYK factor of the original model. The interaction Hamiltonian between the bath and the system is taken as follows

\begin{equation}
    H_{B,S}=\frac{1}{3!}\sum_{ijk}\sum_{\alpha}V_{ijk\alpha}\chi_L^\alpha\psi^i\psi^j\psi^k+\frac{1}{3!}\sum_{ijk}\sum_{\alpha}\tilde{V}_{ijk\alpha}\chi_R^\alpha\tilde{\psi}^i\tilde{\psi}^j\tilde{\psi}^k~.
    \label{eq:bathHamiltonian}
\end{equation}

Again, the couplings $V_{ijk\alpha}$ and $\tilde{V}_{ijk\alpha}$ are taken to be random, with a Gaussian distribution of zero mean, and variance given by
\begin{equation}
    \overline{V_{ijk\alpha}^2}=\overline{\tilde{V}_{ijk\alpha}^2}=\frac{3! V^2}{M^3}~.
\end{equation}
To ensure the system-bath interaction is marginal, each system fermion interacts with three fermions from the bath \cite{Almheiri_2019,MaldaMilekhin}. By choosing $M \gg N$, the backreaction of the system on the bath can be neglected. 

Performing the  disorder average over the couplings $V_{ijk\alpha}$ and $\tilde{V}_{ijk\alpha}$ gives an extra contribution to the effective action \eqref{eq:Seffeuclidean} of the form 
\begin{equation}
    \frac{S[G,\Sigma]_{bath}}{N}= -\frac{V^2}{2}\int d\tau_1 d\tau_2 \left(G_{LL}\left(\tau_1, \tau_2\right)+G_{RR}\left(\tau_1, \tau_2\right)\right)G_B^3\left(\tau_1, \tau_2\right)~.
    \label{eq:bathcontribution}
\end{equation}

In the protocol described above, the coupling $V$ between the system and the bath is turned on at $t=0$, pushing the system out of equilibrium. The sudden coupling breaks time translation invariance, and the evolution of the system must be analyzed using non-equilibrium techniques. The Schwinger-Keldysh (SK) formalism \cite{Kamenev_2011,Haehl:2024pqu} provides a natural framework to do so (see \cite{Eberlein_2017} for its application to the SYK model, and \cite{MaldaMilekhin,Berenguer_2024} for the details on the present case).

\begin{figure}
    \centering
    \begin{tikzpicture}
        % First contour (left side)
        \begin{scope}
            % Axes for first contour
            \draw[->] (-4, 0) -- (4, 0) node[right] {$\mathrm{Re}\,t$};
            \draw[->] (0, -1) -- (0, 1) node[above] {$\mathrm{Im}\,t$};
            
            % Labels
            \node[below left] at (0, 1) {$t_0$};
            \node[above right] at (3, 0) {$+\infty$};
            \node[above left] at (-3, 0) {$-\infty$};
            
            % Contour 1
            \draw[thick,red, ->] (-3, 0.2) -- (-2, 0.2);
            \draw[thick,red, ->] (-2, 0.2) -- (2.05, 0.2);
            \draw[thick,red] (2.05, 0.2) -- (3, 0.2); % Arrow before endpoint
            \draw[thick,red, ->] (3, 0.2) -- (3, -0.1);
            \draw[thick,red] (3, -0.1) -- (3, -0.2); % Arrow before endpoint
            \draw[thick,red, ->] (3, -0.2) -- (1.95, -0.2);
            \draw[thick,red] (1.95, -0.2) -- (0, -0.2);
            \draw[thick,red,->] (0, -0.2) -- (-2.1, -0.2); % Arrow before endpoint
            \draw[thick,red] (-2.1, -0.2) -- (-3, -0.2);
        \end{scope}
        
        % Second contour (right side, aligned with the first)
        \begin{scope}[xshift=7cm] % Shift second diagram to the right
            % Axes for second contour
            \draw[->] (0, 0) -- (4, 0) node[right] {$\mathrm{Re}\,t$};
            \draw[->] (0, -1) -- (0, 1) node[above] {$\mathrm{Im}\,t$};
            
            % Labels
            \node[below left] at (0, 0.5) {$t_0$};
            \node[below left] at (0, -1) {$t_0 - i\beta$};
            \node[above right] at (3, 0) {$+\infty$};
            
            % Contour 1
            \draw[thick,red, ->] (0, 0.2) -- (2.05, 0.2);
            \draw[thick,red] (2.05, 0.2) -- (3, 0.2); % Arrow before endpoint
            \draw[thick,red, ->] (3, 0.2) -- (3, -0.1);
            \draw[thick,red] (3, -0.1) -- (3, -0.2); % Arrow before endpoint
            \draw[thick,red, ->] (3, -0.2) -- (1.95, -0.2);
            \draw[thick,red] (1.95, -0.2) -- (0, -0.2); % Arrow before endpoint
            \draw[thick,red, ->] (0, -0.2) -- (0, -0.6);
            \draw[thick,red] (0, -0.6) -- (0, -1); % Arrow before endpoint
        \end{scope}
    \end{tikzpicture}
    \caption{Contours used in the numerical integration. At $t=t_0$,  the drivings are turned on. Prior to that, the system is at equilibrium. When using the standard predictor-corrector method the initial condition must be explicitly provided (left). When using the NESSi package, the L-shaped contour (right) calculates the initial thermal equilibrium state directly in the Matsubara formalism.}
    \label{fig:Keldyshcontour}
\end{figure}
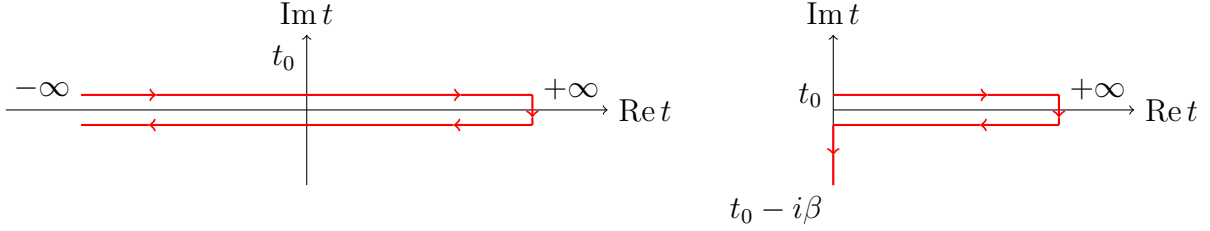

In the SK formalism, observables are computed by evolving the system forwards and backwards in real time along the Keldysh contour $\mathcal{C}$ (see Fig.~\ref{fig:Keldyshcontour}). The real-time counterpart of the Schwinger-Dyson equations \eqref{eq:SDequilibrium}-\eqref{eq:Sigmaeq} are the Kadanoff-Baym equations, which are written in terms of the greater and lesser components of the Green's functions, defined as
\begin{align}
\begin{split}
    G_{ab}^>(t_1,t_2)&=-\frac{i}{N}\sum_j\langle\chi_a^j(t_1)\chi_b^j(t_2)\rangle\\
    G_{ab}^<(t_1,t_2)&=-\frac{i}{N}\sum_j\langle\chi_b^j(t_2)\chi_a^j(t_1)\rangle~,
\end{split}
\end{align}
as well as their retarded, advanced, and Keldysh components,
\begin{align}
\begin{split}
    G_{ab}^R(t_1,t_2)&=\theta(t_1-t_2)\left[G_{ab}^>(t_1,t_2)-G_{ab}^<(t_1,t_2)\right]\\
    G_{ab}^A(t_1,t_2)&=-\theta(t_2-t_1)\left[G_{ab}^>(t_1,t_2)-G_{ab}^<(t_1,t_2)\right]\\
    G_{ab}^K(t_1,t_2)&=G_{ab}^>(t_1,t_2)+G_{ab}^<(t_1,t_2)~,
    \label{eq:retadvkel}
\end{split}
\end{align}
and similarly for the self energies. The final form of the Kadanoff-Baym equations is 
\begin{align}
\begin{split}
    i\partial_{t_1}G_{ab}^>(t_1,t_2)&=i\mu_{ac}(t_1)G_{cb}^>(t_1,t_2)+\int_{-\infty}^{\infty}dt~ \Sigma_{ac}^R(t_1,t)G_{cb}^>(t,t_2)+\int_{-\infty}^{\infty}dt~ \Sigma_{ac}^>(t_1,t)G_{cb}^A(t,t_2)\\
    -i\partial_{t_2}G_{ab}^>(t_1,t_2)&=i G_{ac}^>(t_1,t_2)\mu_{cb}(t_2)+\int_{-\infty}^{\infty}dt~ G_{ac}^R(t_1,t)\Sigma_{cb}^>(t,t_2)+\int_{-\infty}^{\infty}dt ~G_{ac}^>(t_1,t)\Sigma_{cb}^A(t,t_2)~,
\label{eq:KBeqs}
\end{split}
\end{align}
where $\mu_{ab}(t)\equiv(i\sigma_y)_{ab}\mu(t)$, and the self energies are given by
\begin{equation}
    \Sigma_{ab}(t_1,t_2)=-J_a(t_1)J_b(t_2)G_{ab}(t_1,t_2)^3~-V(t_1)V(t_2)G_B(t_1,t_2)^3\delta_{ab}.
\label{eq:SigmaKB}
\end{equation}

The Kadanoff-Baym equations \eqref{eq:KBeqs} and \eqref{eq:SigmaKB} can be solved numerically using a predictor-corrector method on a two-times grid \cite{Eberlein_2017,Bhattacharya_2019,Haldar_2020,Kuhlenkamp_2020,MaldaMilekhin,Zhang_2021,Larzul_2022,Hosseinabadi:2023mid,Berenguer_2024,Guo:2024xio,Jaramillo:2024lul}. However, the non-exponential decay of the Green's functions in the wormhole phase introduces significant challenges when trying to implement this method, particularly in terms of the computational resources required. These challenges can be overcome by using the NESSi library (Non-Equilibrium Systems Simulation \cite{Schuler_2020}).

To track the temperature evolution during the cooling process, we use a variant of the fluctuation-dissipation (FD) relation, which, in equilibrium, states that
\begin{equation}
    \frac{i G^K_{LL}(\omega)}{\rho_{LL}(\omega)}=\tanh\frac{\beta\omega}{2}~,
    \label{eq:FDTeq}
\end{equation}
where $G_{LL}^K(\omega)$ is the Fourier transform of the Keldysh Green's function, defined in \eqref{eq:retadvkel}, and $\rho_{LL}(\omega)$ is the spectral function. Out of equilibrium, the notion of temperature is not well-defined. However, if the cooling process is slow enough, the correlation functions are expected to exhibit near-thermal behavior at a particular effective temperature. One way of defining such temperature is by rotating the time variables into relative time $t$ and average time $\mathcal{T}$, defined as
\begin{equation}
    t=t_1-t_2,~~~~~\mathcal{T}=\frac{t_1+t_2}{2}~,
\end{equation}
and computing the Fourier transform with respect to $t$. For a general function $G(t_1,t_2)\rightarrow G(\mathcal{T},t)$, this yields the Wigner transform $G(\mathcal{T},\omega)$,
\begin{equation}
    G(\mathcal{T},\omega)=\int dt~e^{i\omega t}G(\mathcal{T},t)~.
    \label{eq:Wignertrans}
\end{equation}
Now one can define the $\mathcal{T}$-dependent quantities $G^K(\mathcal{T},\omega)$, $\rho(\mathcal{T},\omega)$, and fit
\begin{equation}
    \frac{i G^K_{LL}(\mathcal{T},\omega)}{\rho_{LL}(\mathcal{T},\omega)}=\tanh\frac{\beta(\mathcal{T})\omega}{2}
    \label{eq:betaeff}
\end{equation}
at each average time $\mathcal{T}$ to obtain a time-dependent effective temperature. By evaluating $\beta(\mathcal{T})$ during the process, the authors in \cite{MaldaMilekhin} found that the system follows the microcanonical equilibrium curve, going through the region of the phase diagram with negative specific heat, associated with the hot wormhole.

If the whole process happens slowly enough, it is reasonable to expect that the configurations of the system during the evolution will correspond to the true (albeit unstable) solutions of the equilibrium Schwinger-Dyson equations \eqref{eq:SDequilibrium}-\eqref{eq:Sigmaeq}, with the replacement $\Sigma_{LL}\rightarrow \Sigma_{LL}+V^2 G_B^3$ (and similarly for $\Sigma_{RR}$), which accounts for the contributions of the bath. We have checked numerically that this is the case. This means that by extracting the $\mathcal{T}$-dependent spectral functions $\rho(\mathcal{T},\omega)$ we have access to all the information regarding these unstable solutions. In particular, we are going to use this approach to extract the chaos exponents. In this way we go beyond the work in \cite{Nosaka_2020}, where these exponents were obtained for the two stable phases.
However, it is important to notice, as mentioned before, that in order to study the system around the unstable points, we must bring it out of equilibrium and let it relax.
For example, the strategy of coupling to a cold bath modifies the equilibrium solutions (see Fig. \ref{fig:phasediag}), meaning the chaos exponents extracted using this method correspond to a different setup than the one originally considered.

For this reason, we are going to compare the results with the ones obtained using the method of \cite{Berenguer_2024}, consisting of injecting energy to the system through a periodic driving of $\mu$, which does not require to modify the initial system. 

The energy is computed as $E(t)=E_L(t)+E_R(t)+E_{int}(t)$, with
\begin{equation}
\frac{E_L}{N}=-\frac{iJ^2}{4}\int_{-\infty}^{t}dt'\Big[G_{LL}^>(t,t')^4-G_{LL}^<(t,t')^4+G_{LR}^>(t,t')^4-G_{LR}^<(t,t')^4\Big]~,
    \label{eq:E_L}    
\end{equation}
and similarly for $E_R$, while the interaction energy is given by $E_{int}(t)=-N\mu(t)G^>_{LR}(t,t)$.

In the next section we compute the chaos exponents of the hot wormhole phase using the two aforementioned approaches.

\section{Hot wormhole phase and chaos exponents}\label{sec:HWchaos}

In the 2-coupled SYK model, the quantum chaotic behavior can be obtained from the connected part of the following 4-point function, which can be written as a two-point function of the Euclidean correlators,
\begin{align}
\begin{split}
    \frac{1}{N^2}\sum_{i,j}\langle \chi_a^i(\tau_1)\chi_b^i(\tau_2)\chi_c^j(\tau_3)\chi_d^j(\tau_4)\rangle&=\langle G_{ab}(\tau_1,\tau_2) G_{cd}(\tau_3,\tau_4)\rangle\\
    &=\frac{\int \mathcal{D}G\mathcal{D}\Sigma~ G_{ab}(\tau_1,\tau_2)G_{cd}(\tau_3,\tau_4)e^{-S_{eff}\left[G,\Sigma\right]}}{\int \mathcal{D}G\mathcal{D}\Sigma~e^{-S_{eff}\left[G,\Sigma\right]}}~.
\end{split}
\end{align}
In the large $N$ limit, this 4-point function admits the expansion
\begin{equation}
    \langle G_{ab}(\tau_1,\tau_2) G_{cd}(\tau_3,\tau_4)\rangle=\overline{G_{ab}(\tau_1,\tau_2)} ~\overline{G_{cd}(\tau_3,\tau_4)}\left[1+\frac{1}{N}\mathcal{F}_{abcd}(\tau_1,\tau_2,\tau_3,\tau_4)+...\right]~,
\end{equation}
and the connected piece can be obtained from the quadratic fluctuations of the effective action $S_{eff}$ around the saddle point solutions, denoted as  $\overline{G_{ab}(\tau_1,\tau_2)}$ \cite{MaldaStanford}.

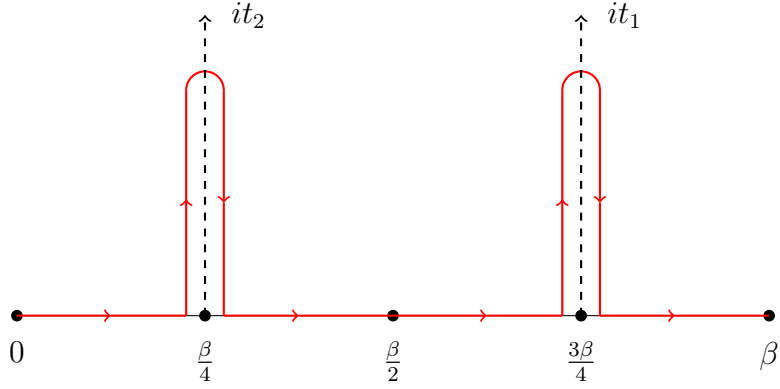
\begin{figure}
    \centering
    \begin{tikzpicture}
        \def\length{5cm}
        \def\eps{0.05*\length}
        \draw[-] (-\length, 0) -- (\length, 0);
        \filldraw[black] (-\length, 0) circle (2pt);
        \filldraw[black] (-\length/2, 0) circle (2pt);
        \filldraw[black] (0, 0) circle (2pt);
        \filldraw[black] (\length/2, 0) circle (2pt);
        \filldraw[black] (\length, 0) circle (2pt);
        % Labels
        \node[below, yshift=-0.2cm] at (-\length, 0) {$0$};
        \node[below, yshift=-0.2cm] at (-\length/2, 0) {$\frac{\beta}{4}$};
        \node[below, yshift=-0.2cm] at (0, 0) {$\frac{\beta}{2}$};
        \node[below, yshift=-0.2cm] at (\length/2, 0) {$\frac{3\beta}{4}$};
        \node[below, yshift=-0.2cm] at (\length, 0) {$\beta$};
        
        % Horizontal red lines
        \draw[thick,red, ->] (-\length, 0) -- (-3*\length/4, 0);
        \draw[thick,red] (-3*\length/4, 0) -- (-\length/2-\eps, 0);

        \draw[thick,red, ->] (-\length/2+\eps, 0) -- (-\length/4, 0);
        \draw[thick,red] (-\length/4, 0) -- (0, 0);

        \draw[thick,red, ->] (0, 0) -- (\length/4, 0);
        \draw[thick,red] (\length/4, 0) -- (\length/2-\eps, 0);

        \draw[thick,red, ->] (\length/2+\eps, 0) -- (3*\length/4, 0);
        \draw[thick,red] (3*\length/4, 0) -- (\length, 0);

        % Vertical red lines
        \draw[thick,red, ->] (-\length/2-\eps, 0) -- (-\length/2-\eps, 0.31*\length);
        \draw[thick,red,] (-\length/2-\eps, 0.3*\length) -- (-\length/2-\eps, 0.6*\length);

        \draw[thick,red, ->] (-\length/2+\eps,  0.6*\length) -- (-\length/2+\eps, 0.3*\length);
        \draw[thick,red] (-\length/2+\eps, 0.3*\length) -- (-\length/2+\eps, 0);

        \draw[thick,red, ->] (\length/2-\eps, 0) -- (\length/2-\eps, 0.31*\length);
        \draw[thick,red,] (\length/2-\eps, 0.3*\length) -- (\length/2-\eps, 0.6*\length);

        \draw[thick,red, ->] (\length/2+\eps,  0.6*\length) -- (\length/2+\eps, 0.3*\length);
        \draw[thick,red] (\length/2+\eps, 0.3*\length) -- (\length/2+\eps, 0);

        % Small semicircles
        
        \draw[thick, red] (-\length/2-\eps, 0.6*\length) arc[start angle=180,end angle=0,radius=\eps];
        \draw[thick, red] (\length/2-\eps, 0.6*\length) arc[start angle=180,end angle=0,radius=\eps];

        % Vertical dashed lines
        \draw[thick,dashed,black, ->] (-\length/2, 0) -- (-\length/2, 0.8*\length) node[right,xshift=0.2cm] {$it_2$};
        \draw[thick,dashed,black, ->] (\length/2, 0) -- (\length/2, 0.8*\length) node[right,xshift=0.2cm] {$it_1$};

    \end{tikzpicture}
    \caption{Usual contour used for the evaluation of the OTOC in the chaos regime.}
    \label{fig:contourLyapunov}
\end{figure}

The Lyapunov exponent is obtained by analytically continuing the time variable, and considering the following ordering of times,
\begin{equation}
    \tau_1=\frac{3\beta}{4}+it_1~,~~~~\tau_2=\frac{\beta}{4}+it_2~,~~~~\tau_3=\frac{\beta}{2}~,~~~~\tau_4=0~,
\end{equation}
with the integration along the contour depicted in Fig.~\ref{fig:contourLyapunov}. The relevant fact is that the  connected piece $\mathcal{F}_{abcd}(\frac{3\beta}{4}+it_1,\frac{\beta}{4}+it_2,\frac{\beta}{2},0)\equiv \mathcal{F}_{abcd}(t_1,t_2)$ satisfies the following kernel equation
(see \cite{Nosaka_2020} for details) 
\begin{equation}
   \mathcal{F}_{abcd}(t_1,t_2)=\sum_{e,f}\int dt_3 dt_4 \, K_{abef}^R(t_1,...,t_4)\mathcal{F}_{efcd}(t_3,t_4)
    \label{eq:kerneleq2c}~
\end{equation}
with the retarded kernel given by
\begin{equation}
    K_{abcd}^R(t_1,...,t_4)=-3J^2G_{ac}^R(t_1-t_3)G_{bd}^R(t_2-t_4)G_{cd}^{W}(t_3-t_4)^2~.
    \label{eq:kernel}
\end{equation}
In \eqref{eq:kernel} $G_{ab}^R(t)$ is a retarded correlator, and $G_{ab}^{W}(t)$ is a Wightman function obtained from the Euclidean correlator as $G_{ab}^{W}(t)=G_{ab}(\frac{\beta}{2}+it)$. Remarkably, Eq. \eqref{eq:kernel} remains valid even in the presence of the bath. The reason is that the way in which we chose to couple each SYK to the bath (Eq. \eqref{eq:bathHamiltonian}) leads to a contribution in the large $N$ effective action that is linear in the 2-point functions of the system (see Eq. \eqref{eq:bathcontribution}). Since \eqref{eq:kernel} is obtained from the quadratic fluctuations of the action, the new terms do not contribute to this order\footnote{This was noticed in \cite{GarciaGarcia_2024} for a similar setup.}.

To solve this equation it is useful to introduce a {\em growing ansatz} of the form
\begin{equation}
    \mathcal{F}_{abcd}(t_1,t_2)=e^{\lambda_L(t_1+t_2)}f_{abcd}(t_1-t_2)~.
    \label{eq:ansatzF}
\end{equation}

For the numerical determination of the Lyapunov exponent, it is convenient to rewrite equation \eqref{eq:kerneleq2c}, with the ansatz \eqref{eq:ansatzF}, in frequency space\footnote{Here $(G_{ef}^{W})^2(\omega)$ refers to the Fourier transform of $G_{ab}^{W}(t)^2$, and not the square of the Fourier transform of $G_{ab}^{W}(t)$.}:
\begin{equation}
    f_{abcd}(\omega)=-3J^2\sum_{e,f}G_{ae}^R(\omega+i\lambda_L/2)G_{bf}^R(-\omega+i\lambda_L/2)\int_{-\infty}^{\infty}d\omega'(G_{ef}^{W})^2(\omega-\omega')f_{efcd}(\omega')~~.
    \label{eq:kerneleqfreq}
\end{equation}
As noted in \cite{Nosaka_2020}, the indices $cd$ do not mix in the equation, so we can take $cd=LL$ and omit this dependence. From now on, $f_{ab}\equiv f_{abLL}$. It is also convenient to express the different correlators in terms of the spectral functions $\rho_{ab}(\omega)$, since these are the outputs of the real-time equilibrium simulations \cite{Berenguer_2024,LantagneHurtubise_2019,Plugge_2020}. With our conventions, the precise relations are
\begin{equation}
    G_{ab}^R(\omega)=-\sigma_{ab}\int d\omega' \frac{\rho_{ab}(\omega')}{\omega-\omega'}~~~~,~~~~~~\sigma=\mathbb{1}+i\sigma_x=\begin{pmatrix}
    1 & i \\
    i & 1
    \end{pmatrix}~
\end{equation}
for the retarded correlator in frequency space, and
\begin{equation}
    G_{ab}(\tau)=\tilde{\sigma}_{ab}\int_{-\infty}^{\infty}d\omega\frac{e^{-\omega\tau}\rho_{ab}(\omega)}{1+e^{-\beta\omega}}~~\rightarrow~~G_{ab}^W(t)=G_{ab}(\beta/2+it)=\tilde{\sigma}_{ab}\int_{-\infty}^{\infty}d\omega\frac{e^{-i \omega t}\rho_{ab}(\omega)}{2\cosh{\frac{\beta\omega}{2}}}~,
\end{equation}
with $\tilde{\sigma}=\mathbb{1}-\sigma_x=\begin{pmatrix}
    1 & -1 \\
    -1 & 1
    \end{pmatrix}$, for the Wightman function. Plugging everything back into \eqref{eq:kerneleqfreq} gives
\begin{equation}
f_{ab}(\omega) =\int d\omega' \sum_{cd}K_{abcd}(\omega,\omega') f_{cd}(\omega')~, \label{eq:fabomega}
\end{equation}
with
\begin{align}
\begin{split}
    K_{abcd}(\omega,\omega')=-\frac{3J^2}{4} \sigma_{ac}\sigma_{bd}\int d\omega_1\frac{\rho_{ac}(\omega_1)}{\omega-\omega_1+i\lambda_L/2}&\int d\omega_2\frac{\rho_{bd}(\omega_2)}{-\omega-\omega_2+i\lambda_L/2}\\ \rule{0mm}{7mm}
    \times&  \int d\omega_3\, \frac{\rho_{cd}(\omega_3)\rho_{cd}(\omega-\omega'-\omega_3)}{\cosh\displaystyle\frac{\beta\omega_3}{2}\cosh\frac{\beta(\omega-\omega'-\omega_3)}{2}}~.
    \label{eq:fabomega2}
\end{split}
\end{align}

\iffalse
\begin{align}
\begin{split}
    f_{ab}(\omega)=-\frac{3J^2}{4}\sum_{c,d}\sigma_{ac}\sigma_{bd}\int d\omega_1\frac{\rho_{ac}(\omega_1)}{\omega-\omega_1+i\lambda_L/2}&\int d\omega_2\frac{\rho_{bd}(\omega_2)}{-\omega-\omega_2+i\lambda_L/2}\\
    \times&\int d\omega'\int d\omega_3\frac{\rho_{cd}(\omega_3)\rho_{cd}(\omega-\omega'-\omega_3)}{\cosh\frac{\beta\omega_3}{2}\cosh\frac{\beta(\omega-\omega'-\omega_3)}{2}}f_{cd}(\omega')~,
    \label{eq:fabomega}
\end{split}
\end{align}

 has the form

\begin{equation}
    \begin{pmatrix}
    f_{LL} \\
    f_{LR} \\
    f_{RL} \\
    f_{RR}
    \end{pmatrix}=
    \begin{pmatrix}
    K_{LLLL} & K_{LLLR} & K_{LLRL} & K_{LLRR} \\
    K_{LRLL} & ... &  & \\
    K_{RLLL} &  & ... & \\
    K_{RRLL} &  &  & ...
    \end{pmatrix}
    \begin{pmatrix}
    f_{LL} \\
    f_{LR} \\
    f_{RL} \\
    f_{RR}
    \end{pmatrix}~,
\end{equation}
\fi

Eq. \eqref{eq:fabomega} tells us that $f_{ab}(\omega)$ is an eigenfunction of the kernel matrix $K$ given in \eqref{eq:fabomega2} with eigenvalue 1. The Lyapunov exponent corresponds to the value of $\lambda_L$ for which the largest eigenvalue of $K$ crosses 1.

\subsection{Chaos exponents: wormhole formation}

\begin{figure}
    \centering
    % Left column: two stacked subfigures
    \begin{minipage}[t]{0.3\textwidth}
        \centering
        \begin{subfigure}[t]{\textwidth}
            \centering
            \includegraphics[width=\textwidth]{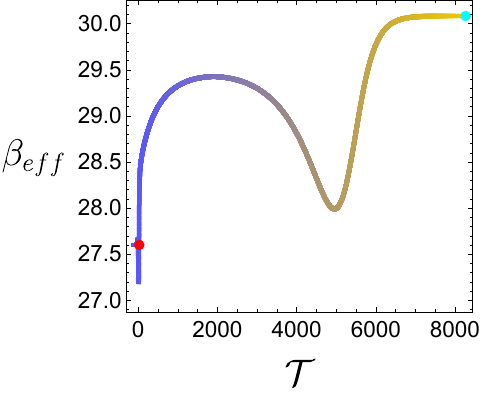}
        \end{subfigure}
        \vspace{1cm}
        \begin{subfigure}[b]{\textwidth}
            \centering
            \includegraphics[width=\textwidth]{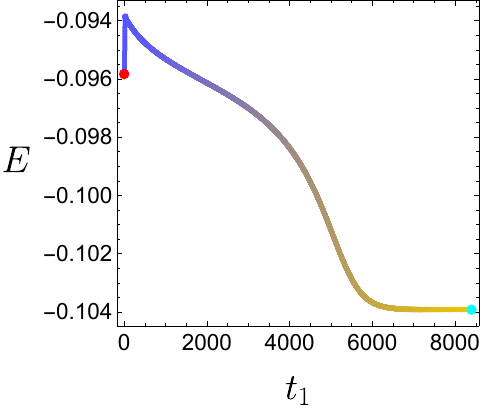}
        \end{subfigure}
    \end{minipage}
    \hspace{0.02\textwidth} % Space between the columns
    % Right column: single subfigure
    \begin{minipage}[c]{0.45\textwidth}
        \centering
        \begin{subfigure}[t]{\textwidth}
            \centering
            \includegraphics[width=\textwidth]{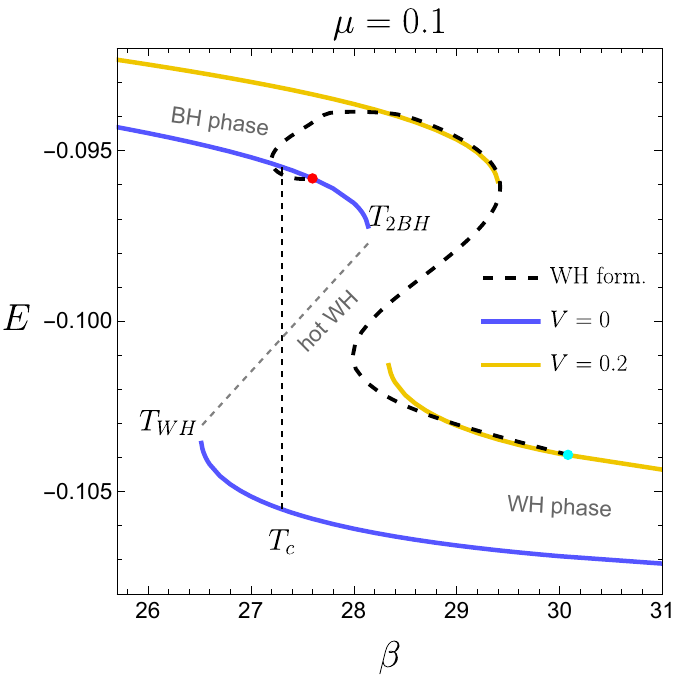}
        \end{subfigure}
    \end{minipage}
    \caption{Wormhole formation in real time \cite{MaldaMilekhin}. We show the cooling process that allows us to transition from the black hole to the wormhole phase by turning on a coupling to a cold bath. The system explores the hot wormhole solutions of the new phase diagram. On the left we show the inverse effective temperature and the energy during the process. The two quantities are combined into the black dashed line in the right plot. The color gradient on the left plots stresses the fact that we begin in the original closed system, but we end up in the system containing the bath.}
    \label{fig:WHformation}
\end{figure}

We analyze first the protocol in which the cooling process is done by means of a cold bath. Throughout the simulations, we fix the coupling strengths of the SYK factors to $J=J_B=1$ and the interaction between the two sides to $\mu=0.1$. In this protocol, we initialize the system in the black hole phase, at a temperature slightly higher than $T_{2BH}$. We have chosen $\beta_i=27.6$. At $t=0$, a coupling to a cold bath at $\beta_B=30$, with interaction strength $V=0.2$, is turned on. We turn it on linearly between $t=0$ and $t=20$ to avoid abrupt changes in the parameters, which would render additional, undesired non-equilibrium effects. The numerical results in Fig. \ref{fig:WHformation} show that the equilibration process brings the system from the black hole to the wormhole solutions, going through a region where the temperature increases. The latter is identified with the negative specific heat solutions corresponding to the hot wormhole, in agreement with \cite{MaldaMilekhin}. We also overlap the time-dependent energy and temperature with the equilibrium phase diagram.

From Fig.~\ref{fig:WHformation} we observe that, as expected, introducing the bath alters the hot wormhole solutions making them differ from those initially sought. In Section \ref{sec:FloquetSYK}, we will revisit this issue using a different non-equilibrium protocol that avoids modifying the original problem.

\begin{figure}[h!]
    \centering
    \includegraphics[width=0.45\textwidth]{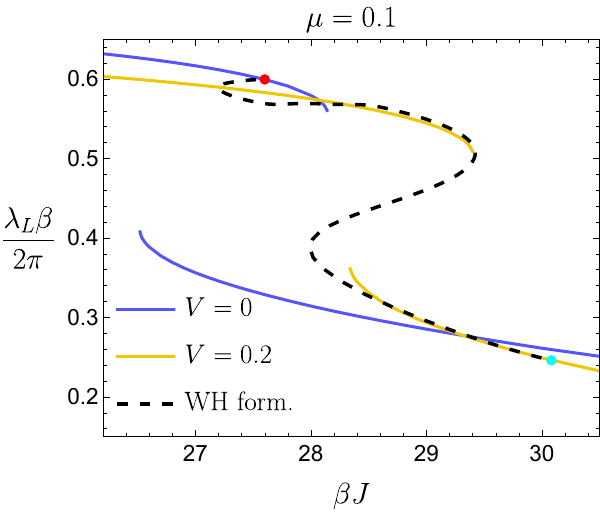}
    \caption{Chaos exponents of the hot wormhole phase obtained at each average time $\mathcal{T}$ during the cooling process.}
    \label{fig:WHformchaos}
\end{figure}

We can ignore this for now, and proceed with the determination of the chaos exponents, using \eqref{eq:fabomega}. Fig.~\ref{fig:WHformchaos} shows, in blue,  the chaos exponents of the BH and WH phases of the original model, which were already obtained in \cite{Nosaka_2020}. In yellow, the ones corresponding to the system coupled to the baths in equilibrium. In dashed we show the chaos exponents obtained from our non-equilibrium simulation, where the spectral function and temperature at each average time $\mathcal{T}$ are obtained from the non-equilibrium Green's functions as in \eqref{eq:Wignertrans}-\eqref{eq:betaeff}. We observe how, similarly as it happened with the energy, the Lyapunov exponent of the hot wormhole solutions interpolates quite smoothly between the two phases. This  matching was already conjectured in Fig. 18 of \cite{Nosaka_2020}. As mentioned above, the caveat of this protocol is that the system for which we are exploring the unstable phase is not the original one, but rather the one in contact with the cold baths.

\begin{figure}[h!]
    \centering
    \includegraphics[width=0.49\textwidth]{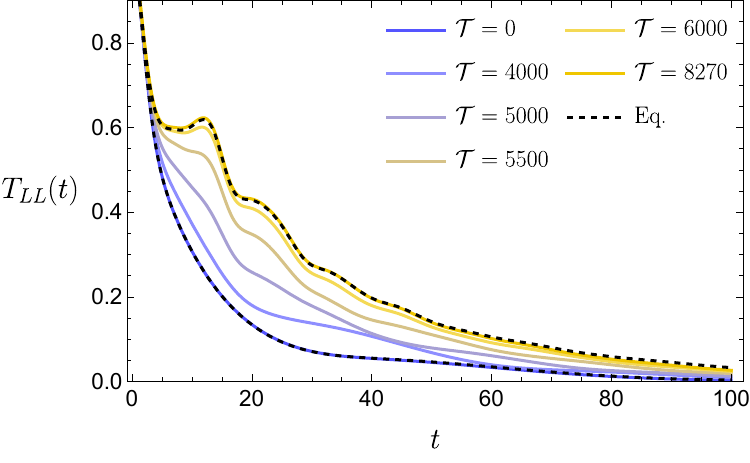}
    \hspace{0.05cm}
    \includegraphics[width=0.49\textwidth]{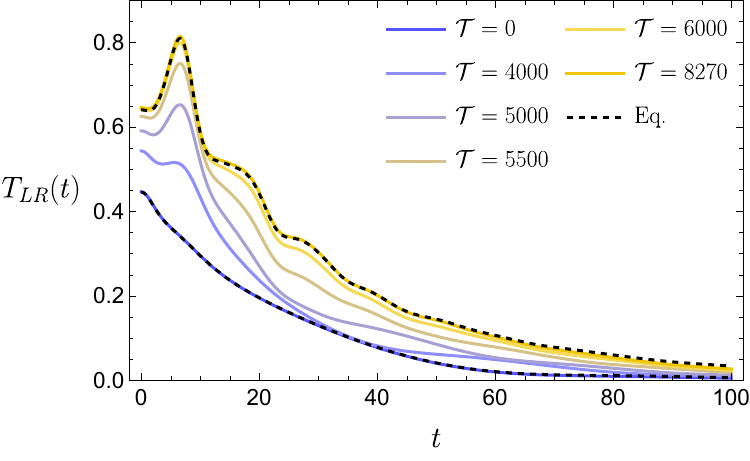}
    \caption{Wormhole formation in real time. By coupling the system, initially in a black hole solution ($\mathcal{T}=0$), to a cold bath, a wormhole solution can be reached once the equilibrium with the bath is achieved ($\mathcal{T}=8270$ in our simulations). The dashed lines show the transmission coefficients of the equilibrium solutions for $\beta=27.6$ and $\beta=30$ (initial and end points in Fig. \ref{fig:WHformation}). The coupling to  the baths reduces the observed revivals in the wormhole phase in comparison with the isolated case (see \cite{LantagneHurtubise_2019,Berenguer_2024} and Fig.~\ref{fig:TLLLRFloquet} below).}
    \label{fig:TLLLRWHformation}
\end{figure}

Once $\beta_{eff}(\mathcal{T})$ and $\rho_{ab}(\omega,\mathcal{T})$ have been computed, as demanded for the calculation of the chaos exponents, they can further be also used to extract the pseudo-equilibrium Green's functions $G_{ab}^>(\mathcal{T},t)$, $G_{ab}^<(\mathcal{T},t)$, and from them, the transmission amplitudes, $T_{ab}(\mathcal{T},t)=2\abs{G_{ab}^>(\mathcal{T},t)}$ \cite{Berenguer_2024}. Plotting these quantities allows to see visually the process of wormhole formation studied in \cite{MaldaMilekhin}, as we show in Fig.~\ref{fig:TLLLRWHformation}.

\subsection{Chaos exponents: Floquet SYK wormholes} \label{sec:FloquetSYK}

\begin{figure}
    \begin{minipage}[t]{0.35\textwidth}
        \centering
        \begin{subfigure}[t]{\textwidth}
            \centering
            \includegraphics[width=\textwidth]{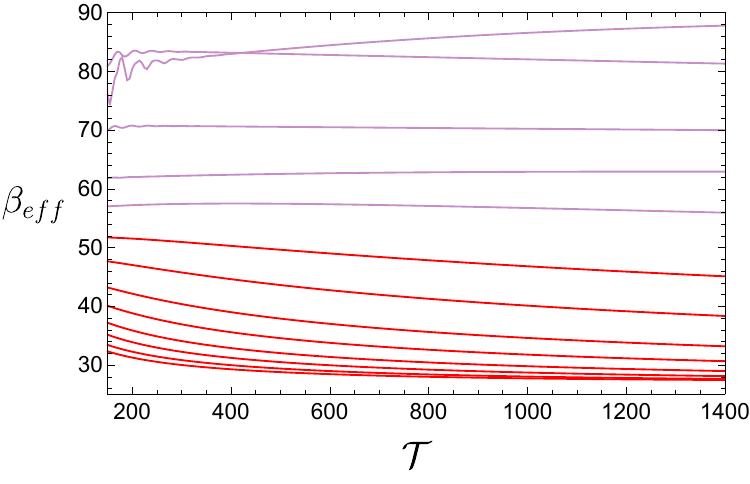}
            \caption{}
            \label{sfig:betaeffnonthermal}
            \vspace{0.1cm}
        \end{subfigure}
        \begin{subfigure}[b]{\textwidth}
            \centering
            \includegraphics[width=\textwidth]{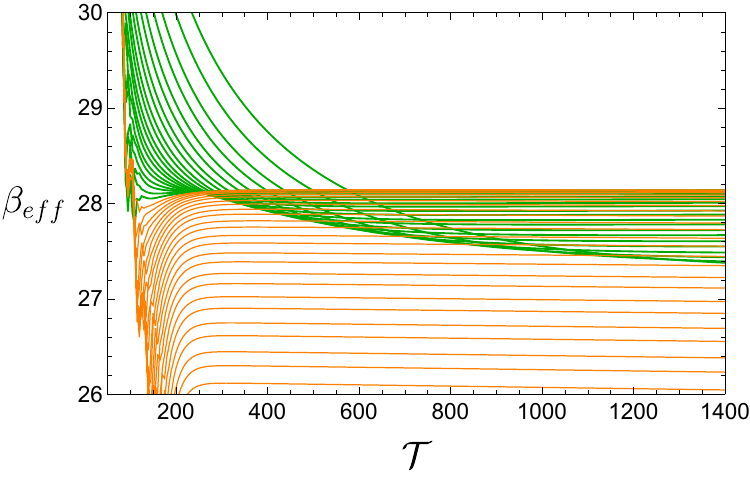}
            \caption{}
            \label{sfig:betaeffthermal}
        \end{subfigure}
    \end{minipage}
    \hspace{0.01\textwidth}
    \centering
    \vspace{0.3cm}
    \begin{minipage}[c]{0.55\textwidth}
        \vspace{0.7cm}
        \centering
        \begin{subfigure}[t]{\textwidth}
            \centering
            \vspace{1cm}
            \includegraphics[width=\textwidth]{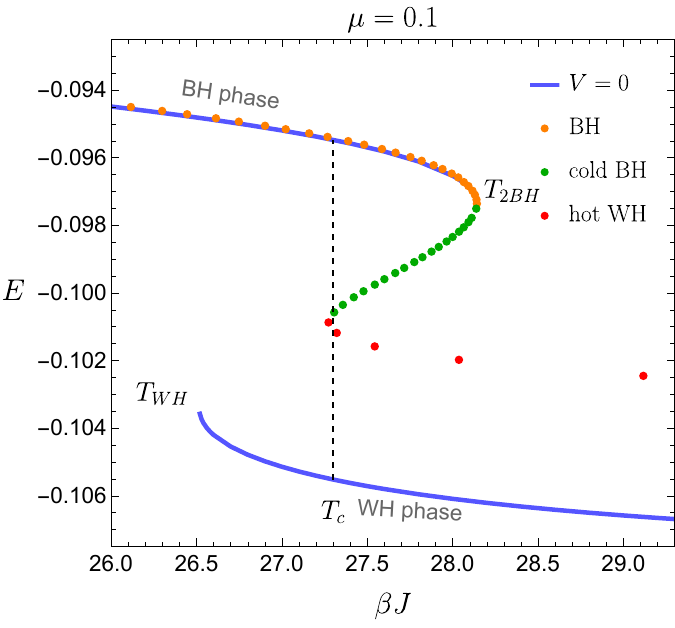}
            \caption{}
            \label{sfig:Floquetphasediagram}
        \end{subfigure}
    \end{minipage}
    \begin{subfigure}[c]{0.49\textwidth}
    \includegraphics[width=0.49\textwidth]{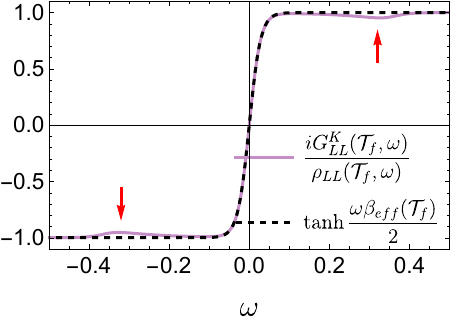}
    \includegraphics[width=0.49\textwidth]{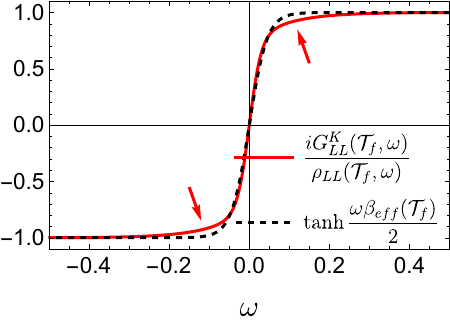}
    \caption{}
    \label{sfig:tanhnonthermal}
    \end{subfigure}
    \begin{subfigure}[c]{0.49\textwidth}
    \includegraphics[width=0.49\textwidth]{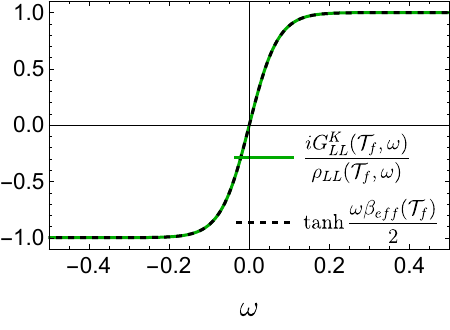}
    \includegraphics[width=0.49\textwidth]{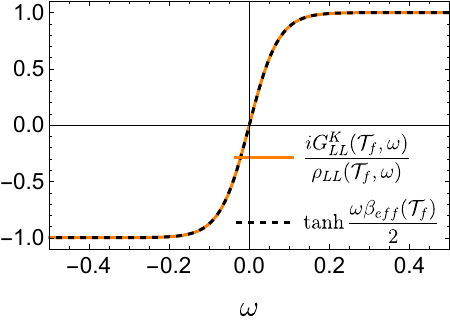}
    \caption{}
    \label{sfig:tanhthermal}
    \end{subfigure}
    \caption{Figs. \ref{sfig:betaeffnonthermal}, \ref{sfig:betaeffthermal}: Effective inverse temperature for the non-thermal ($n\in \left[1,13\right]$ in Fig. \ref{sfig:betaeffnonthermal}) and thermal ($n\in \left[14,69\right]$ in Fig. \ref{sfig:betaeffthermal}) solutions. Only the asymptotic values in \ref{sfig:betaeffthermal} can be interpreted as proper temperatures since only there the FD is satisfied exactly. Fig. \ref{sfig:Floquetphasediagram}: we locate these solutions in the phase diagram and observe that the green thermal ones belong to the hot wormhole phase. The presence of the (red) non-thermal solutions in the diagram is explained in Section \ref{sec:HotWHColdBH}. Fig. \ref{sfig:tanhnonthermal}: FD relation at large $\mathcal{T}$ for $n=3, 6$ (non-thermal). Fig. \ref{sfig:tanhthermal}: FD relation at large $\mathcal{T}$ for $n=25, 50$ (thermal).}
    \label{fig:Floquetbetasdiagram}
\end{figure}

We consider again the initial system, namely, two coupled SYK models, with the hamiltonian given in \eqref{eq:Hamiltonian}. In this part of the analysis we effectively explore the microcanonical ensemble by injecting controlled amounts of energy. We do so by periodically driving the parameter $\mu$, following the protocol described in \cite{Berenguer_2024}, which we review here. The initial equilibrium state is a wormhole with $\mu=0.1$ and $\beta_i=30$, which is very close to the transition temperature, $\beta_c\sim 27.3$. At $t=0$, we turn on the driving as $\mu(t)=\mu\left(1+a\sin\left(\Omega t\right)\right)$, with $a=0.1$ and $\Omega=1.0$, which corresponds to a resonant frequency of the wormhole phase (see \cite{Berenguer_2024}), thereby injecting energy into the system at an exponential rate. After $n$ half cycles, the driving is stopped at $t_{\text{stop}}^n=\frac{\pi}{\Omega}n$, and $\mu$ returns to its static initial value, $\mu=0.1$. As a result, the energy of the system increases for $t<t_{\text{stop}}^n$ and stabilizes at a constant value $E_f$ beyond that point. Asymptotically, the system is then expected to reach a state, which may or may not be a thermal equilibrium state. 

We repeat the simulation for increasing values of $n$, which allows to characterize the thermalization under different energy injections. In each simulation, we analyze the asymptotic behavior of the system. Our numerical capabilities are limited to $t_{max}=1500$, which is enough to characterize the thermalization in most cases. The results are shown in Fig.~\ref{fig:Floquetbetasdiagram}, where two main scenarios have to be distinguished: for $n<14$, the system does not fully thermalize within our simulation time. By this we mean that the fluctuation-dissipation relation \eqref{eq:betaeff} does not hold exactly. As a result, the values of $\beta_{eff}(\mathcal{T})$ obtained from the fits are not meaningful, and therefore we cannot use \eqref{eq:fabomega} to determine the chaos exponents for these solutions. We show the evolution of the (meaningless) effective inverse temperature of the non-thermal solutions in Fig. \ref{sfig:betaeffnonthermal}. Each line corresponds to a different $t_{\text{stop}}^n$ ($n=1,2,3...$). The fact that the results of the fit to \eqref{eq:betaeff} become $\mathcal{T}$-independent for the purple lines could suggest that the effect of the driving has been a cooling of the system to a much larger $\beta$ than the initial one. We want to stress that this is not exactly the case, since the FD relation (although $\mathcal{T}$-independent), is not satisfied, as we show in Fig. \ref{sfig:tanhnonthermal} for two particular choices of $n$. The colors purple and red denote two distinct behaviors within the non-thermal solutions, whose details are relegated to Section \ref{sec:HotWHColdBH}, where we will also explore whether they could thermalize over longer timescales.

\begin{figure}[h!]
    \centering
    \includegraphics[width=0.45\textwidth]{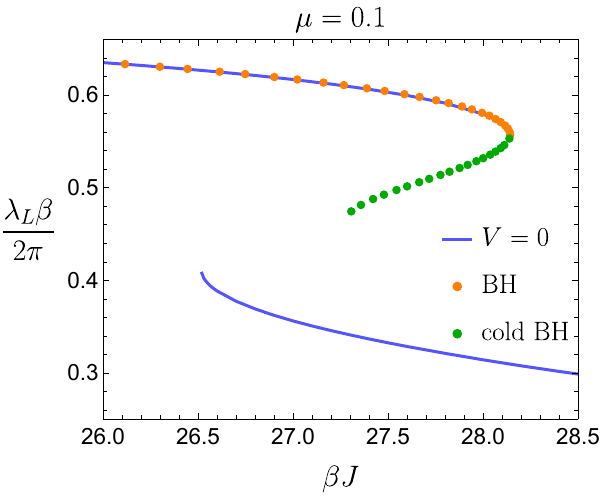}
    \caption{Chaos exponents of the final equilibrium solutions that thermalize within our simulation time. With our protocol we are able to explore approximately half of the unstable region of the phase diagram.}
    \label{fig:Floquetdiagramchaos}
\end{figure}

Conversely, for $n\geq14$, the system relaxes to a thermal state after a finite time. Coincidentally, when this change of behavior occurs, $\beta_{eff}\left(\mathcal{T} \to \infty\right)\approx \beta_{\text{c}}$. We show the thermal solutions in Fig. \ref{sfig:betaeffthermal}. Again, in Fig. \ref{sfig:tanhthermal} we check the FD relation, which holds for these solutions and allows us to obtain the final inverse temperature of the system. In \ref{sfig:Floquetphasediagram} we locate the thermal solutions in the equilibrium phase diagram, where we find that these solutions belong to the hot wormhole phase. Again, the two colors in the set of thermal solutions signal a change of behavior in the way the effective inverse temperature approaches its final value. Comparing Figs. \ref{sfig:betaeffthermal} and \ref{sfig:Floquetphasediagram} we notice this change in concavity/convexity coincides with the moment the system ceases to stabilize in a hot wormhole solution, and reaches the black hole phase. 

In general, we observe that our driving protocol fails to capture all possible states within the coexistence region, instead it only converges to points along the upper half segment of the hot wormhole phase. In Section \ref{sec:Schwarzian} we will use the Schwarzian approximation to show that this is what we should expect. In other words, we will show that for small energy injections, the system does not transition to the hot wormhole phase, but instead it explores excited states of the cold wormhole.

In all these cases (thermal solutions of the hot wormhole and black hole solutions), it is possible to compute the chaos exponents. Remarkably, these correspond to the ones we were seeking at first, which do not require the addition of a bath. The results are shown in Fig.~\ref{fig:Floquetdiagramchaos}.

\subsubsection{The last dance: a mixture of both approaches}
A natural question is whether the results obtained above reflect intrinsic properties of the model or are merely artifacts of the chosen driving protocol. To address this, we consider an alternative protocol that incorporates a combination of the two methods above. In this case, starting from a black hole solution, the system is coupled to a cold bath for a finite time before being decoupled as it transitions through the (modified) hot wormhole phase. The goal is to determine whether energy extraction, rather than injection, allows the system to stabilize within the previously unobserved region of the hot wormhole branch. Once the bath is turned off, the system goes back to its original form, and we expect the final equilibrium states to align with those of the unperturbed model, in the lines of the analysis above.
\begin{figure}[h!]
    \begin{minipage}[t]{0.45\textwidth}
        \centering
        \begin{subfigure}[t]{\textwidth}
            \centering
            \includegraphics[width=\textwidth]{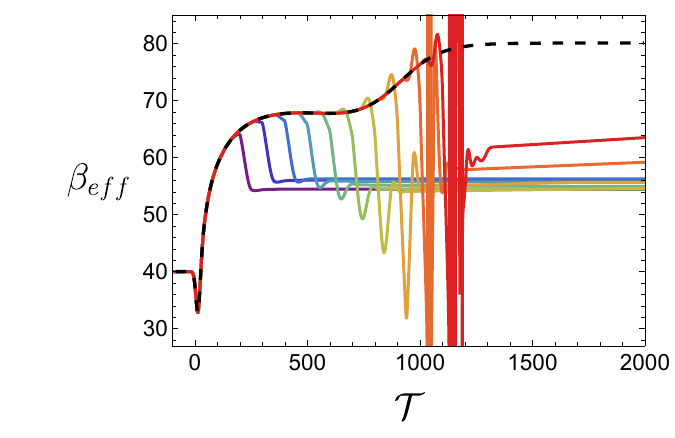}
            \vspace{-0.4cm}
        \end{subfigure}
        \begin{subfigure}[b]{\textwidth}
            \centering
            \includegraphics[width=\textwidth]{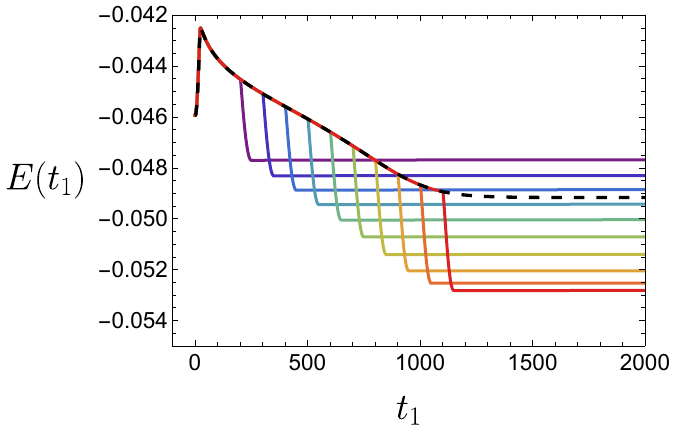}
        \end{subfigure}
    \end{minipage}
    \hspace{0.01\textwidth}
    \centering
    \vspace{0.3cm}
    \begin{minipage}[c]{0.5\textwidth}
        \vspace{0.7cm}
        \centering
        \begin{subfigure}[t]{\textwidth}
            \centering
            \vspace{1cm}
            \includegraphics[width=\textwidth]{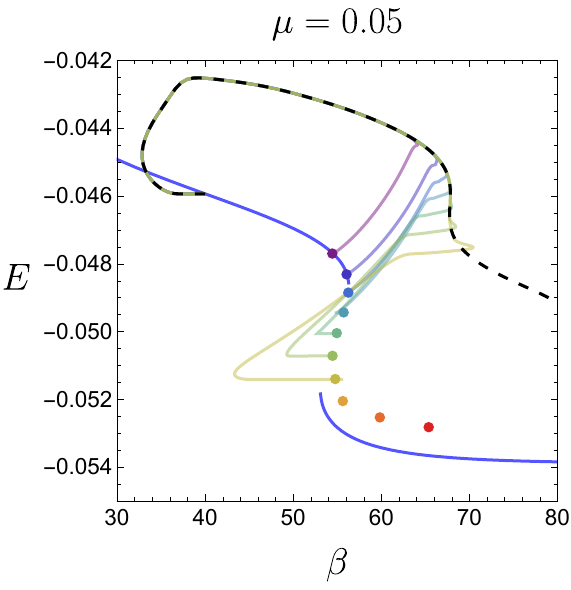}
        \end{subfigure}
    \end{minipage}
    \begin{subfigure}[c]{\textwidth}
    \centering
    \includegraphics[width=0.3\textwidth]{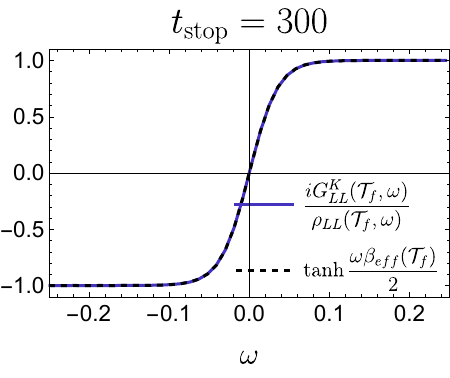}
    \includegraphics[width=0.3\textwidth]{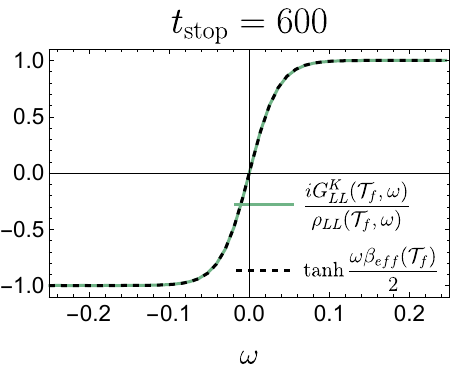}
    \includegraphics[width=0.3\textwidth]{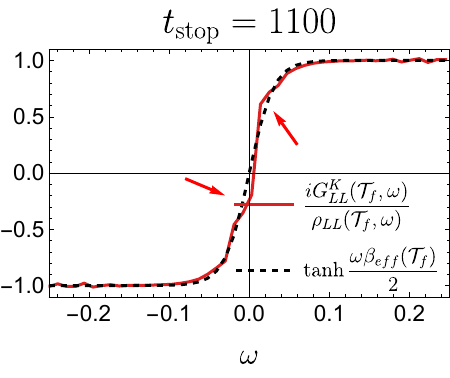}
    \label{sfig:tanhthermalstopbath}
    \end{subfigure}
    \caption{The figure displays the time evolution of the inverse effective temperature and energy, along with their trajectory in the $E$-$\beta$ phase diagram (right plot). For visual clarity, the trajectory lines for the last three points, where $\beta$ exhibits strong oscillations, have been omitted. $t_{\text{stop}}$ increases from purple to red, taking values $t_{\text{stop}}=200,300,...,1100$. The lower panels present the fluctuation-dissipation relation for three representative cases, illustrating the distinct behaviors observed.}
    \label{fig:stopbathresults}
\end{figure}

By turning off the coupling to the bath at different $t_{\text{stop}}$, we can extract controlled amounts of energy. The results\footnote{For this analysis we are using different parameters than in the previous section. In particular, we are using the same parameters as in \cite{MaldaMilekhin}, where $J=J_B=0.5$, $\mu=0.05$, and $V=0.2$. The initial inverse temperature is $\beta_i=40$, and the bath is at $\beta_B=80$.}, shown in Fig. \ref{fig:stopbathresults}, reveal the same qualitative behavior as in previous cases—namely, this protocol captures only the equilibrium solutions corresponding to the upper segment of the hot wormhole branch. In these solutions, the fluctuation-dissipation relation holds, and equilibration to the final state occurs almost instantaneously. However, for larger energy extractions, the system fails to thermalize, leading to a time-dependent effective temperature that lacks physical meaning.

This provides further evidence that the structure of the unstable region of the phase diagram is more intricate than initially thought, and obtaining the complete set of hot wormhole solutions is non-trivial. Motivated by this, the next section examines the structure of this phase in greater detail.

\section{Hot wormholes vs cold black holes}\label{sec:HotWHColdBH}

\begin{comment}
The two stable phases identified in \cite{MaldaQi} were named the "two black hole phase" for sufficiently high temperatures $T_{BH} < T$, and the "cold wormhole" phase for sufficiently low temperatures $T < T_{WH}$. In the intermediate coexistence region, where $T_{BH} < T < T_{WH}$, a canonically unstable conjectured phase was referred to as the "hot wormhole." \cite{MaldaMilekhin}. In this section, we argue that within this intermediate region, there may exist a substructure that can be categorized into "cold black holes" and "hot wormhole."

The periodic driving protocol employed to construct states in this region involves first exciting the system and subsequently allowing it to relax to a thermal state at a specific temperature. As explained in Section \ref{sec:FloquetSYK}, this method does not capture all possible states within the coexistence region, instead it only converges to points along the upper segment of what has traditionally been referred to as the "hot wormhole phase." The precise alignment of the phase transition (vertical line) with the point where the asymptotic $\beta_{eff}$ starts increasing is unlikely to be coincidental. Furthermore, it is beyond this point, that the fluctuation-dissipation (FD) relations are satisfied exactly, indicating that thermalization is inhibited prior to this point. These observations suggest that the so-called "hot wormhole phase" may itself contain a substructure comprising the "hot wormhole" and "cold black hole" states. 

\end{comment}

\begin{figure}[!ht]
    \centering
    \includegraphics[width=0.49\textwidth]{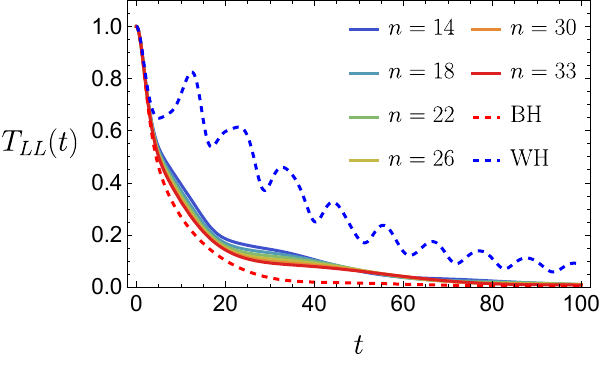}
    \hspace{0.05cm}
    \includegraphics[width=0.49\textwidth]{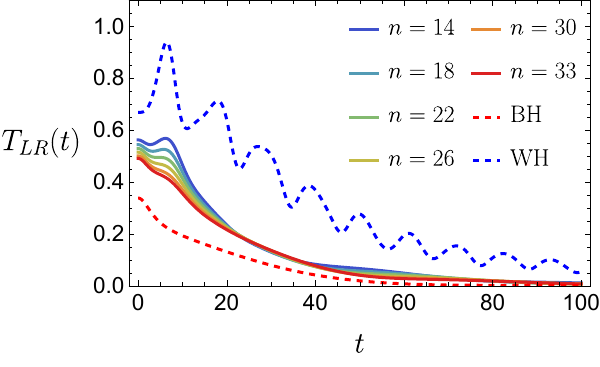}
    \caption{Transmission amplitudes of four different final states within the unstable phase. We show typical profiles of the amplitudes for the black hole phase (dashed red for $\beta=20$), and the wormhole phase (dashed blue for $\beta=30$).}
    \label{fig:TLLLRFloquet}
\end{figure}

In this section we put the focus on the structure of the unstable region of the phase diagram. We begin by showing the behavior of the thermal solutions from the previous section corresponding to the unstable branch (green solutions in Figs.~\ref{fig:Floquetbetasdiagram} and~\ref{fig:Floquetdiagramchaos}, with $n \in \left[14,33\right]$). In Fig.~\ref{fig:TLLLRFloquet} we show the transmission amplitudes $T_{LL}(t)$ and $T_{LR}(t)$ for some of them. We also show, for comparison, the typical profiles of solutions belonging to the black hole phase (dashed red lines), and the wormhole phase (dashed blue lines). The transmission amplitudes of the new solutions show that as we go deeper in the unstable phase (lower $n$), a peak in the transmission amplitude $T_{LR}$ emerges. However, their overall behavior is closer to a black hole than to a wormhole.

For the non-thermal solutions, i.e. purple and red curves in Fig. \ref{sfig:betaeffnonthermal}, with $n\in [1,13]$, we can distinguish two classes. For the purple curves,  the results of the fits to the (not precisely satisfied) FD relation are approximately $\mathcal{T}$-independent. In the next subsection we interpret these as excited states of the stable wormhole phase. 

The other solutions (in red in  Fig. \ref{sfig:betaeffnonthermal})  pose greater challenges for interpretation. They show a slowly varying (although meaningless) effective temperature. The fits to the FD relation improve as $\mathcal{T}$ increases, indicating a potential thermalization at late times. In order to extract the endpoint of this thermalization, we fit the red solutions to an exponential of the form
\begin{equation}
    \beta_{eff}(\mathcal{T})=Ae^{-\gamma \mathcal{T}}+\beta_{\infty}~,
\end{equation}
with $\beta_{\infty}=\lim_{\mathcal{T}\rightarrow \infty}\beta_{eff}(\mathcal{T})$. The reasonable expectation is that these solutions would fill the interpolating segment in the hot wormhole phase between from $\beta_c\sim 27.3$ to the wormhole phase. Surprisingly, the results of the exponential fits give $\beta_\infty>\beta_c$, i.e. the red dots in Fig. \ref{sfig:Floquetphasediagram}. These {\em almost} thermalized solutions are suspicious from the point of view of the phase diagram. Their true existence as thermal equilibrium solutions  would imply   that the system suffers a decrease in temperature after a very small injection of energy. If the fits to the FD relation improve with time but never become exact, these solutions have a nature similar to the purple non-thermal solutions and the temperature is not well-defined. If, on the contrary, the FD relation is satisfied at asymptotically late times, the mechanism that governs this thermalization is unknown. This effect is very similar to the observation that the sharp revival oscillations in the wormhole phase have an envelope that decays as a power-law in time \cite{Plugge_2020}, suggesting the possibility that the two phenomena could be related.

In addition to this, we observe that the turning point that signals the transition between thermal (green) and non-thermal (red) solutions is aligned with the temperature of first order Hawking-Page like  transition between the black hole phase and the wormhole phase, at $\beta_c\sim 27.3$. These observations, together with the behavior of the transition amplitudes shown in  Fig. \ref{fig:TLLLRFloquet} suggest that the thermalized purple solutions should be rather called "cold black holes". The  {\em true} hot wormholes, that would eventually fill the rest of the segment joining the two stable phases, remain inaccessible to our driving protocol.

\subsection{Schwarzian analysis}\label{sec:Schwarzian}
In this section we want to provide qualitative arguments for why we can only access a sub-region of the unstable phase with our protocol. In other words, why the system only thermalizes for energy injections above a certain threshold. Although for the values of the parameters used in the simulations, the Schwarzian approximation does not allow for precise quantitative comparisons\footnote{For smaller values of $\mu$, where the Schwarzian approximation is valid, the phase transition occurs at higher $\beta$. In that regime, the characteristic decay times of the Green's functions drastically increases and so do the computational resources needed for non-equilibrium simulations.}, the qualitative behavior is similar \cite{MaldaMilekhin} and it allows us to gain some intuition.

\begin{figure}[ht]
    \centering
    \includegraphics[width=0.5\textwidth]{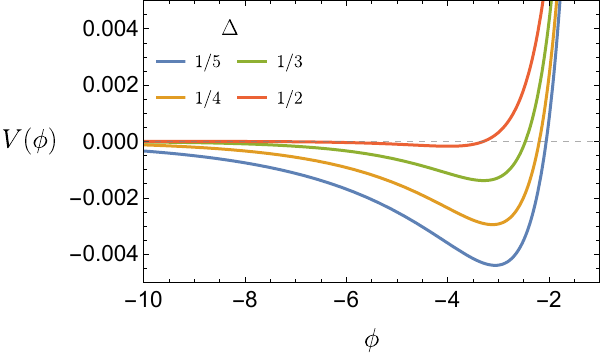}
    \caption{Shape of the potential \eqref{eq:potential} for different values of $\Delta$. The case $\Delta=1/2$ can be solved analytically.}
    \label{fig:potential}
\end{figure}

In the low energy limit $\mu\ll J$ the model is effectively governed by the following Schwarzian action \cite{MaldaQi}
\begin{equation}
    S=N\int du \left[-\frac{\alpha_S}{\mathcal{J}}\left(\left\{\tan\frac{t_l(u)}{2},u\right\}+\left\{\tan\frac{t_r(u)}{2},u\right\}\right)+\mu\frac{c_\Delta}{(2\mathcal{J})^{2\Delta}}\left(\frac{t'_l(u)t'_r(u)}{\cos^2{\frac{t_l(u)-t_r(u)}{2}}}\right)^\Delta\right]~,
    \label{eq:Schwaction}
\end{equation}
where $2\mathcal{J}^2=J^2$, $c_\Delta=\frac{1}{2}\left[(1-2\Delta)\frac{\tan \pi\Delta}{\pi\Delta}\right]^\Delta$, and $\alpha_S$ is a numerical constant \cite{MaldaStanford}.
It can be shown \cite{MaldaQi} that the solutions can always be gauge transformed into the form $t_l(u)=t_r(u)\equiv t(u)$, and the action reduces to
\begin{equation}
    S=-N~\frac{2\alpha_S}{\mathcal{J}}\int du \left[\left\{\tan\frac{t(u)}{2},u\right\}-\tilde{\mu}~t'(u)^{2\Delta}\right],
    \label{eq:Schwactiontprime}
\end{equation}
with $\tilde{\mu}\equiv \mu\frac{\mathcal{J}c_\Delta}{2\alpha_S(2\mathcal{J})^{2\Delta}}$. In terms of a new field $\phi(u)$, defined as $t'(u)=e^{\phi(u)}$, the equations of motion of \eqref{eq:Schwactiontprime} are
\begin{equation}
    \phi''(u)+e^{2\phi(u)}-2\Delta\tilde{\mu}e^{2\Delta\phi(u)}=0~,
    \label{eq:EoMphi}
\end{equation}
which correspond to the motion of a particle under the influence of the potential
\begin{equation}
    V(\phi)=\frac{1}{2}e^{2\phi}-\tilde{\mu}e^{2\Delta\phi}~,
    \label{eq:potential}
\end{equation}
which we show in Fig.~\ref{fig:potential}.

The simplest solution to \eqref{eq:EoMphi} has the linear form $t(u)=t'u$, where $t'$ is a constant. This solution corresponds to the configuration in which the particle sits at the minimum of the potential. We want to consider all the other solutions to the equation of motion \eqref{eq:EoMphi}. By looking at the potential we see that these will be bounded and oscillating solutions when $E<0$, but they are unbounded when $E>0$, with the energy given by
\begin{equation}
    E=\frac{2\alpha_S}{\mathcal{J}}\left(\frac{1}{2}\phi'^2+\frac{1}{2}e^{2\phi}-\tilde{\mu}e^{2\Delta\phi}\right).
    \label{eq:energySchwarz}
\end{equation}

\begin{figure}[ht]
    \centering
    \includegraphics[width=0.5\textwidth]{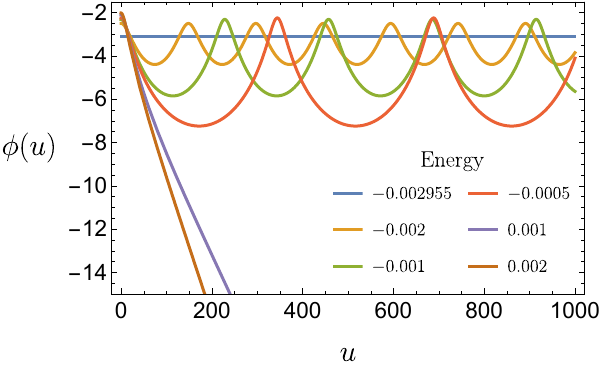}
    \caption{Typical bounded ($E<0$) and unbounded ($E\ge 0$) solutions of the potential \eqref{eq:potential}. The constant blue line corresponds to the solution sitting at the minimum.}
    \label{fig:solutionsE}
\end{figure}

The solutions for arbitrary values of the energy can be found analytically when $\Delta=1/2$ \cite{Dhar:2018pii}. For the present case ($\Delta=1/4$) they have to be obtained numerically. Typical profiles of the solutions are shown in Fig.~\ref{fig:solutionsE}.

Let's focus on the constant solution at the minimum of the potential. We aim to demonstrate that we can bring it to an oscillating solution with a higher energy by turning on the driving of the coupling $\tilde{\mu}$. For this, we change $\tilde{\mu}\rightarrow\tilde{\mu}(u)$ in the action \eqref{eq:Schwaction}. The same change translates into the equation of motion \eqref{eq:EoMphi}.

\begin{figure}[ht]
    \centering
    \includegraphics[width=0.6\textwidth]{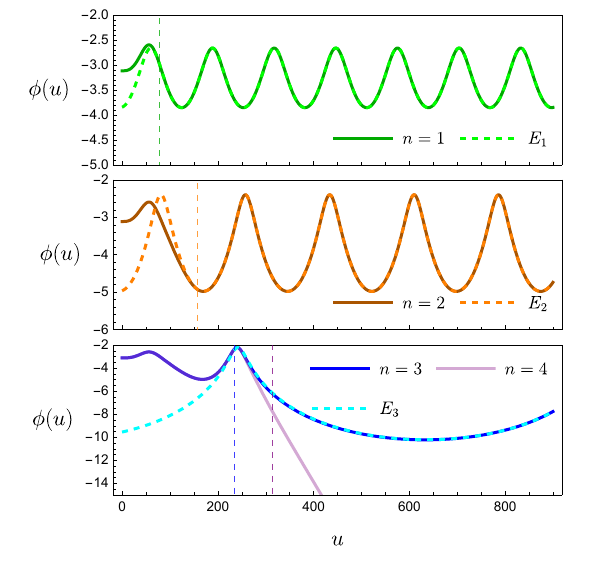}
    \caption{Solid lines: numerical solutions of the equation of motion \eqref{eq:EoMphi} for different $u_{\text{stop}}$ (marked as vertical lines). In particular, $n=1,2,3,4$. For $n=1,2,3$, the injection of energy turns out to be small enough to remain inside the potential. The final energy is evaluated as \eqref{eq:energySchwarz} and the solution of the equation of motion with the corresponding final energy is overlaid in dashed lines. The case $n=4$ corresponds to a driving that has increased the energy to a positive value, leading to an unbounded solution.}
    \label{fig:solutionsdriving}
\end{figure}

We can again solve numerically this equation for the different drivings $\tilde{\mu}(u)=\tilde{\mu}(1+a\sin\Omega u)$, where we turn on the driving at $u=0$, but we stop it at $u_{\text{stop}}=\frac{\pi}{\Omega}n$. We solve it for different $u_{\text{stop}}$, with $n=1,2,3,...$. In all the cases, we choose as initial conditions, $\phi(0)=\phi_0$, $\phi'(0)=0$, which corresponds to the solution at the minimum. The results (Fig.~\ref{fig:solutionsdriving}) show that once the driving is turned on, $\phi(u)$ begins to oscillate. As soon as we turn off the driving (marked as dashed, vertical lines) two distinct behaviors appear: in all cases, the energy has increased, and depending on the magnitude of this increase, the solution remains bounded, or becomes unbounded. We evaluate the final energy of these solutions and check that they correspond to the oscillating solutions of Fig.~\ref{fig:solutionsE} with the corresponding energy (dashed lines).

This proves that, by periodically driving the coupling $\tilde{\mu}$, we excite the solutions with higher energy in this potential. These are the non-thermal solutions that we find for small injections of energy, which we understand as excited solutions of the stable wormhole phase. For higher energies, the particle escapes from the potential, leading to the thermal solutions.

\subsubsection{Matter contribution}
The previous Schwarzian analysis offers a simplified yet insightful framework for understanding the dynamics of the system, revealing two classes of behavior: oscillating solutions that remain perpetually trapped within the potential, corresponding to the purple lines in Fig. \ref{sfig:betaeffnonthermal}, and unbounded solutions that escape to infinity, transitioning the system into the black hole phase (orange lines in Fig. \ref{sfig:betaeffthermal}). However, the Schwarzian picture fails to capture the full range of observed behaviors, leaving two critical types of solutions unexplained.

First, there are numerical solutions that reach the hot wormhole phase (green lines in Fig. \ref{sfig:betaeffthermal}). In order to capture the existence of this unstable phase in the Schwarzian picture, one needs to include in the free energy the contributions of the matter fields in the throat. At low enough temperature this can be approximated by the contribution of the lightest excitation of the theory, consisting of a fermion with mass $\Delta=1/4$, with $F\approx -\frac{1}{\beta}e^{-t'\beta/4}$  \cite{MaldaMilekhin}. Such contribution introduces a correction to the Schwarzian potential \eqref{eq:potential}, resulting in the following change
\begin{equation}
    V(\phi)\rightarrow \tilde{V}(\phi,\beta)=\frac{1}{2}e^{2\phi}-\tilde{\mu}e^{\phi/2}-\frac{\mathcal{J}}{2\alpha_S\beta}e^{-\frac{\beta}{4}e^{\phi}}~.
\label{eq:effectivepotential}
\end{equation}

This potential retains the minimum of the original potential, but for low enough $\beta$ (high enough $T$) it develops a local maximum, which corresponds precisely to the unstable hot wormhole solution. For a sufficiently high energy injection, the particle may overcome the barrier and roll down to $\phi\to -\infty$, to meet the black hole solution. According to this picture, there should be a single (and very fine-tuned) value of the injected energy for which the particle stays at the maximum. However, this feature does not seem to occur for a concrete energy injection. The existence of the green solutions in Figs. \ref{sfig:betaeffthermal} and \ref{sfig:Floquetphasediagram} indicates that there is a intermediate range of energy injections for which the effective particle will end up sitting for a long time on top of the local maximum, in agreement with the metastable character of the cold black hole solutions. This precise fine-tuning may come from the fact that a consistent treatment of \eqref{eq:effectivepotential} should involve a $\beta_{eff}(\mu(t))$ dependence, since the driving induces a change of the effective temperature. This dynamical backreaction of the potential in response to the driving may give rise to an attractor mechanism that, for a range of such energy injections, makes the maximum of the potential and the rest point of the particle to meet with high precision. While this fine-tuning might appear suspicious, it is ultimately dictated by the Schwinger-Dyson equations, which admit a single solution for a given energy.

\iffalse
 Despite this modification, the Schwarzian picture faces significant challenges. Notably, the shape of the potential exhibits a strong $\beta$-dependence, which undermines its consistency with a dynamical driving of $\tilde{\mu}$. Indeed,  when the coupling $\tilde{\mu}$ is periodically driven, the effective temperature $\beta_{eff}$ also evolves, causing the potential to change dynamically.  Intriguingly, for the green solutions, the shape of the potential adjusts precisely to ensure that, at asymptotically late times, its maximum precisely aligns with the injected energy, allowing us to "sit" at the maximum. While this fine-tuning might appear questionable, it is ultimately dictated by the Schwinger-Dyson equations, which admit a single solution for a given energy.
\fi

\begin{figure}[ht]
    \centering
    \includegraphics[width=0.49\textwidth]{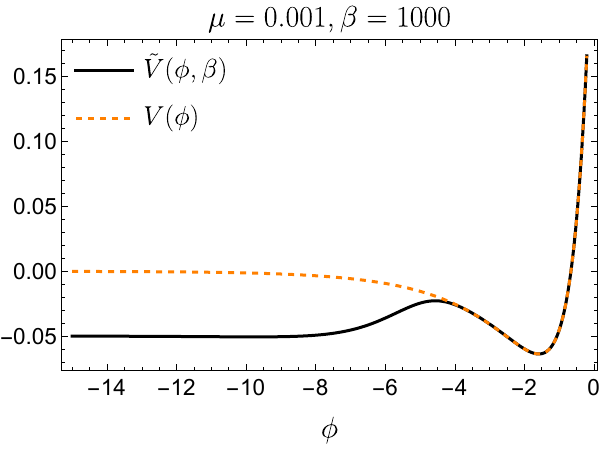}
    \includegraphics[width=0.49\textwidth]{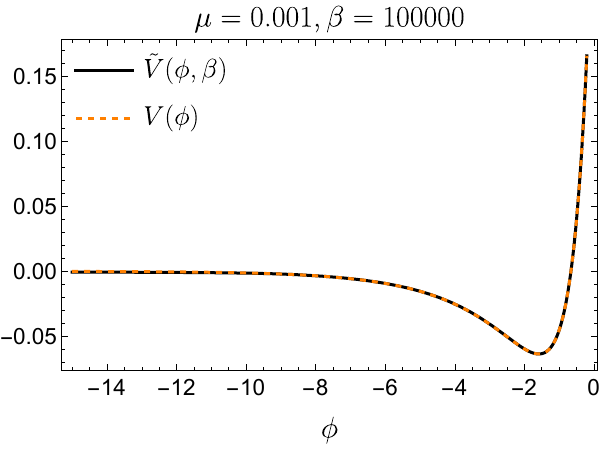}
    \caption{Original (orange, dashed) and modified (black) potential for two different temperatures.}
    \label{fig:newpotential}
\end{figure}

As we have seen, the Schwarzian description is not devoid of  significant challenges.
It also does not fully capture  the dynamics of the red solutions in Fig. \ref{sfig:betaeffnonthermal}. As explained in the previous section, these solutions show a slowly varying (albeit not well defined) effective temperature, with an improvement of the fits to the FD relation as $\mathcal{T}$ increases, suggesting the possibility of an eventual thermalization at late times. The picture of the Schwarzian potential does not provide a mechanism for this as, for trapped solutions, the oscillations should persist indefinitely.

\section{Conclusions}\label{sec:conclusions}

In this work, we analyzed the unstable “hot wormhole” phase of the two-coupled SYK model, which is inaccessible through equilibrium simulations. Using two distinct non-equilibrium protocols, we explored this phase dynamically. The first protocol involves coupling the system to a cold bath, enabling a quasi-static cooling process that transitions the system from a black hole to a wormhole configuration. In the second protocol we drive the $LR$ coupling, $\mu$, periodically in time, injecting energy directly into the system.

The chaotic dynamics of the hot wormhole phase is characterized in both cases through the computation of the Lyapunov exponent. While the first method modifies the equilibrium properties due to the coupling to the bath, the second protocol probes the hot wormhole phase of the original model. Remarkably, driving $\mu$ reveals previously unobserved non-equilibrium behaviors, including thermal and non-thermal states. The emergence of these states is confirmed again with a protocol that combines the two previous ones, in which we couple the system to a cold bath, but the coupling is turned off after certain time. 

A partial understanding of these phenomena is provided by a Schwarzian analysis, though it does not capture the full range of observed behaviors. This suggests a richer structure within the unstable phase, which warrants further investigation. A complete reformulation of the problem in the microcanonical ensemble, where the negative heat phase is stable, would be the most appropriate framework to answer these open questions.

The two-coupled SYK model is far from being maximally chaotic, and tighter chaos bounds proposed in the literature \cite{Avdoshkin:2019trj, Gu:2021xaj} merit comparison with our numerical results. The numerical techniques developed for this work and for \cite{Berenguer_2024} are easily adaptable to other setups with equilibrium and non-equilibrium protocols. A related system was proposed in \cite{Garcia-Garcia:2020ttf, Garcia-Garcia:2022zmo}, introducing complex couplings in the SYK factors, which led to a Euclidean-to-Lorentzian wormhole transition.

Finally, a large-$q$ analysis of the model is also promising, given the existence of analytical results in this limit \cite{MaldaStanford, MaldaQi}. This could provide greater control over the transition region and a deeper understanding of the non-thermal solutions. Additionally, broader studies of quantum dynamics and scrambling, including Krylov complexity, may offer new insights into the interplay between chaos, complexity, and gravity.

\section*{Acknowledgements}
We would like to thank warmly Felix Haehl, Alexey Milekhin and Adrián Sánchez-Garrido for long and illuminating discussions. 

Simulations in this work were made possible through the access granted by the Galician Supercomputing Center (CESGA) to its supercomputing infrastructure. The supercomputer FinisTerrae III and its permanent data storage system have been funded by the Spanish Ministry of Science and Innovation, the Galician Government and the European Regional Development Fund (ERDF).
This work has received financial support from the Xunta de Galicia (CIGUS Network of Research Centres and grant ED431C-2021/14), the European Union, the María de Maeztu grant CEX2023-001318-M funded by MICIU/AEI/10.13039/501100011033 and the Spanish Research State Agency (grants PID2020-114157GB-I00 and PID2023-152148NB-I00).

The work of MB has been funded by Xunta de Galicia through the Programa de axudas \'a etapa predoutoral da Xunta de Galicia (Conseller\'ia de Cultura, Educaci\'on e Universidade) and the grant 2023-PG083 with reference code ED431F 2023/19. The work of J.SS. was supported by MCIN/AEI/10.13039/501100011033 and FSE+ with the grant PRE2022-102163.

\appendix
%\section{Appendix}

\bibliographystyle{JHEP}
\bibliography{chaosHW}

\end{document}